\documentclass[acmtog]{acmart}

\usepackage[dvipsnames]{xcolor}
\usepackage{overpic}
\usepackage{algorithm}
\usepackage{algpseudocode}
\usepackage{float}
\usepackage{xspace}
\usepackage{booktabs} %
\usepackage{soul}
\usepackage{amsmath}
\usepackage{xcolor}
\usepackage{tcolorbox}
\tcbuselibrary{breakable, listings}

\acmJournal{TOG}

\newif\ifdraft
\draftfalse

\ifdraft

\newcommand{\dale}[1]{{\color{blue}[\textbf{Dale:} #1]}}
\newcommand{\jacob}[1]{{\color{red}[\textbf{Jacob:} #1]}}
\newcommand{\rana}[1]{{\color{purple}[\textbf{Rana:} #1]}}
\newcommand{\oded}[1]{{\color{orange}[\textbf{Oded:} #1]}}
\newcommand{\nad}[1]{{\color{green}[\textbf{Nam Anh:} #1]}}
\newcommand{\amir}[1]{{\color{cyan}[\textbf{Amir:} #1]}}

\else
\newcommand{\dale}[1]{}
\newcommand{\jacob}[1]{}
\newcommand{\rana}[1]{}
\newcommand{\oded}[1]{}
\newcommand{\nad}[1]{}
\newcommand{\amir}[1]{}

\fi

\newtcblisting{promptbox}[1][]{
    breakable,
    colback=gray!8,
    colframe=gray!8,
    boxrule=0pt,
    arc=3pt,
    left=8pt, right=8pt,
    top=8pt, bottom=8pt,
    listing only,
    listing options={
        basicstyle=\small\sffamily,
        breaklines=true,
        columns=fullflexible,
        breakindent=0pt,
    },
    #1                        
}

\newcommand{\ourmethod}{CrossLift}

\makeatletter
\DeclareRobustCommand\onedot{\futurelet\@let@token\@onedot}
\def\@onedot{\ifx\@let@token.\else.\null\fi\xspace}

\def\ie{\emph{i.e}\onedot}

\makeatother

\let\titleold\title
\renewcommand{\title}[1]{\titleold{#1}\newcommand{\thetitle}{#1}}

\def\maketitleappendix
   {
   \newpage
       \twocolumn[
        \centering
        \Large
        \textbf{\thetitle}\\
        \vspace{0.5em}Appendix \\
        \vspace{1.0em}
       ] %
   }

\AtEndPreamble{
    \usepackage[capitalize]{cleveref}
    \crefname{section}{Sec.}{Secs.}
    \Crefname{section}{Section}{Sections}
    \Crefname{table}{Table}{Tables}
    \crefname{table}{Tab.}{Tabs.}
}

\usepackage{bm}

\usepackage{mathtools}

\let\originalleft\left
\let\originalright\right
\renewcommand{\left}{\mathopen{}\mathclose\bgroup\originalleft}
\renewcommand{\right}{\aftergroup\egroup\originalright}

\DeclareMathAlphabet{\pazocal}{OMS}{zplm}{m}{n}
\SetMathAlphabet\pazocal{bold}{OMS}{zplm}{bx}{n}

\DeclareMathSymbol{\shortminus}{\mathbin}{AMSa}{"39}

\newcommand{\gap}{\phantom{.}}

\newcommand{\nl}{\gap\\}

\newcommand{\degree}{^\circ}

\newenvironment{Indent}{\nl\begin{adjustwidth}{2em}{0em}}{\end{adjustwidth}\nl}

\newcommand\restr[2]{{%
  \left.\kern-\nulldelimiterspace %
  #1 %
  \vphantom{\big|} %
  \right|_{#2} %
  }}

\newcommand{\ldsym}{$\left.\mathstrut\right)$}%
\newlength{\ldwidth}

\newcommand{\longDivide}[2]%
{\settowidth{\ldwidth}{\ldsym}
#2\,\raisebox{1.5pt}{\ldsym}\hspace*{-.65\ldwidth}\overline{
\mathstrut\hspace*{.35\ldwidth}\ #1}}

\newenvironment{lang}{\setlist{nolistsep}
\begin{enumerate}[$\mid$]}{\end{enumerate}}

\begin{document}
\title{Look Both Ways Before You Cross: Lifting Cross Fields From 2D Visual Priors}

\author{Dale Decatur}
\affiliation{%
 \institution{University of Chicago}
 \country{USA}
}
\author{Jacob Serfaty}
\affiliation{%
 \institution{University of Chicago}
 \country{USA}
}
\author{Oded Stein}
\affiliation{%
 \institution{University of Southern California}
 \country{USA}
}
\affiliation{%
 \institution{Technion}
 \country{Israel}
}

\author{Amir Vaxman}
\affiliation{%
 \institution{University of Edinburgh}
 \country{United Kingdom}
}
\author{Rana Hanocka}
\affiliation{%
 \institution{University of Chicago}
 \country{USA}
}

\begin{abstract}
We present \ourmethod{}, a technique for computing cross fields on meshes guided by visual features in images. We leverage powerful text-to-image priors that are capable of synthesizing images of feature-aligned quad meshes in 2D. We extract this signal as explicit per-pixel directions in the 2D images, which we then back-project to the mesh surface. We aggregate these candidate surface directions by performing two smooth interpolations on the mesh surface (first within each view and second across multiple views). We propose custom confidence-based weights for the candidate directions in each interpolation that allow us to resolve conflicts between candidates on the same face and smoothly interpolate our field to occluded faces.\dale{if we don't have an ablation on these weights, is it ok to leave in abstract or should we remove this previous sentence?}\oded{I think it's fine. We'll have one in the final version.} Our method is modular and can be used with many different 2D visual priors. We show additional applications to texture-aligned quad meshing as well as interactive cross-field design using coarse, user-drawn lines as signal. We demonstrate the effectiveness of \ourmethod{} on a diverse set of both organic and mechanical shapes and produce quad meshes that exhibit superior semantic alignment as compared to existing methods.
\amir{General comment: make sure you use the correct TOG format. It's similar, but maybe some parameters, like Submission ID etc., are different.}
\end{abstract}

\ccsdesc[500]{Computing methodologies~Mesh geometry models}

\keywords{Quad Meshing, Cross Fields, Direction Fields, Visual Loss}

\begin{teaserfigure}
    \centering
    \newcommand{\pl}{-1}
    \includegraphics[width=\textwidth, trim=20 0 20 40]{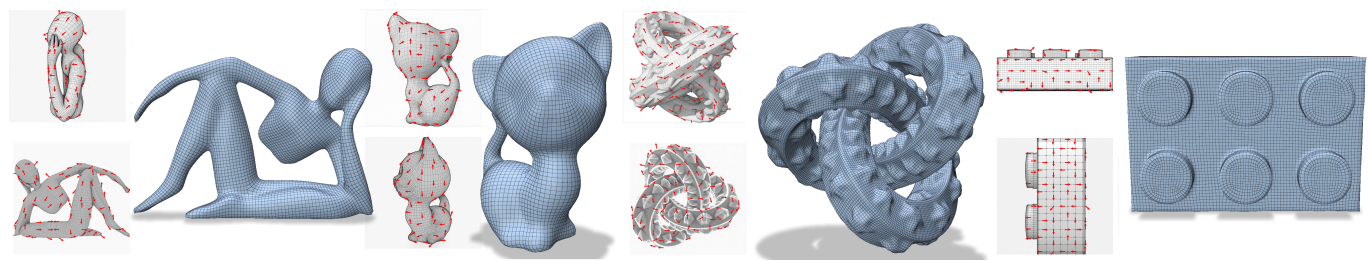}
    \captionof{figure}{\ourmethod{} produces feature-aligned quad meshes (\textcolor{blue}{blue}) using directions (\textcolor{red}{red}) extracted from image-space quad alignment priors (\textcolor{gray}{gray}). These priors encode both semantic and geometric information about the shape allowing our resulting quad meshes to capture meaningful semantic features (e.g., the facial features of the cat) while adhering to local geometric cues such as the sharp ridges of the knot and curved edges of the Lego brick studs.}
    \label{fig:teaser}
\end{teaserfigure}

\maketitle

\section{Introduction}

Quadrilateral (quad) meshes are ubiquitous in 3D modeling. They provide an intuitive, user-friendly representation as their natural structure lends itself to downstream tasks such as animation, parameterization, and editing~\cite{bommes2013quad}. A key advantage of quad meshes is the ability to align them to semantic and geometric directions on the shape, such as arms of a person or ears of a kitten (see \cref{fig:teaser}). This concept of feature alignment closely corresponds to the way artists sketch, making quad meshes the de-facto representation for 3D modeling \cite{bommes2012state, girshick2000line}.

\begin{figure*}[t]
    \centering
    \includegraphics[width=\linewidth]{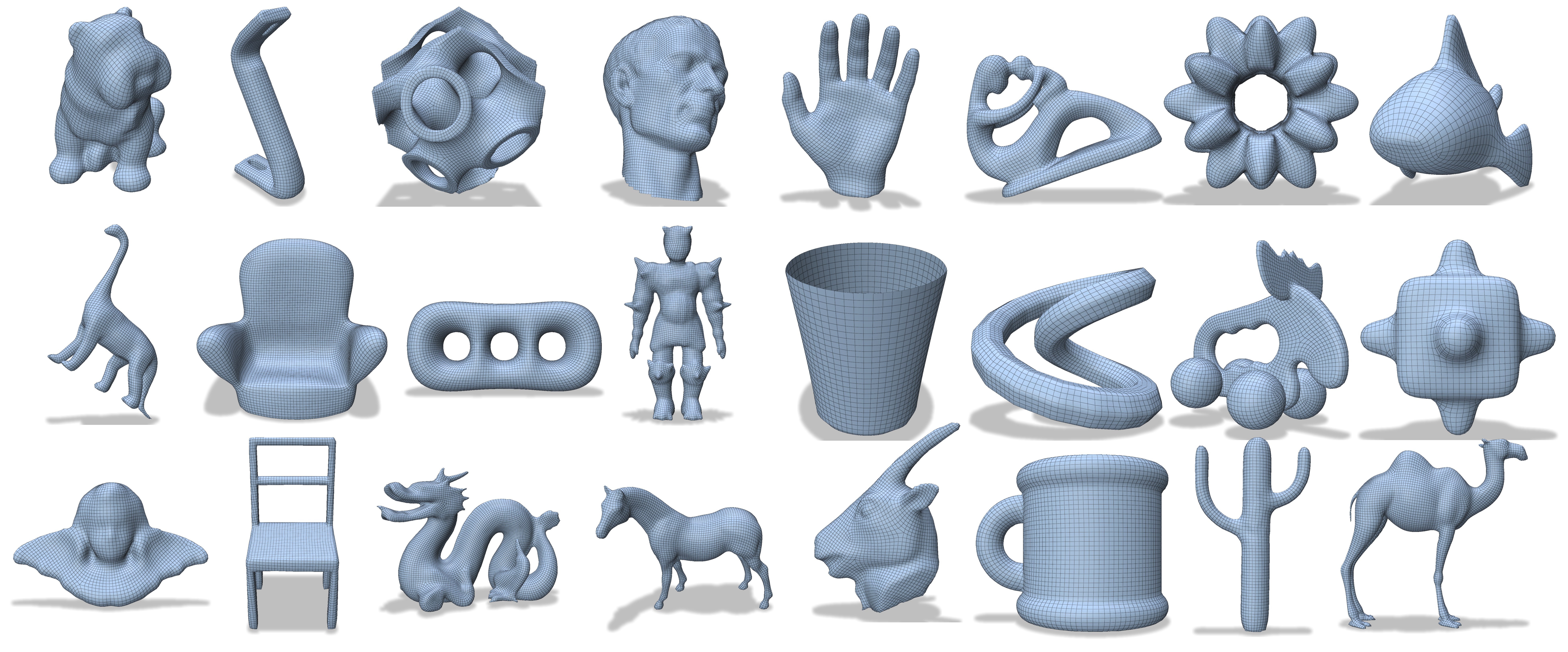}
    \vspace{-8mm}
    \caption{\textbf{Gallery}. Our method produces desirable quads for a variety of shapes, both organic and manufactured. 
    The quad lines do not just follow geometric features:
    our method generates quads whose edges follow semantic feature lines of the shapes through our use of text-to-image priors and without any 3D interaction.}
    \label{fig:gallery}
\end{figure*}

Aligning quads to feature directions is non-trivial; artists and 3D modelers often design the individual quads by hand on the original 3D surface to create a desirable quad layout~\cite{BlenderStudio2026}. Not only is this process time-intensive and tedious, but it requires significant training and expertise to perform effectively. The challenge of manual quad design has driven research on automated quad meshing algorithms~\cite{10.1111:cgf.13498, 10.1145/3450626.3459941, NeuFrameQ-25}. Many automated methods opt to re-mesh existing triangulated meshes to obtain a quad mesh. A common approach is to produce a $4$-\textit{RoSy} (four-way rotationally symmetric) ``cross'' field over the surface of the input mesh that encodes the alignment of the quads at each face. This cross-field can then be integrated to obtain a parametrization, the integer isolines of which can be extracted to form the quad edges. Existing methods for quad meshing typically rely on mesh geometry to align their cross fields~\cite{dong2025neurcrossneuralapproachcomputing, dong2025crossgen}. 

While geometric properties are often a good proxy for feature alignment, they are typically local signals that do not take into account the global properties of the shape, a necessary component for robust feature alignment. In contrast, our method is guided by 2D image priors which contain both geometric cues and global semantic understanding. For example, on the Lego brick in \cref{fig:teaser}, our method captures the geometric radial symmetry at the edges of the raised circular studs and aligns quads accordingly. Then in the flat regions in between the studs that do not contain as well defined geometric signal, our method relies on the semantic cues to align quads to the global notion of the primary axes of the shape (up/down and left/right in this case). Our visual guidance also ensures our method is robust to geometric noise (see \cref{fig:wrinkles}) and even allows us to align to visual features with no geometric components (see \cref{fig:texture-alignment}).

We present \ourmethod{}, a method for designing feature-aligned cross fields using 2D visual signals as guidance.
\amir{I'm missing a sentence on why 2D visual signals are a great idea (low-effort, plugging in existing models, etc.})\dale{I think I address this in the following paragraph, but not sure if maybe I should move it up?}
Our method leverages priors from text-to-image generative models trained on vast amounts of 2D data to produce multi-view images that capture both geometric and semantic quad alignment for the input mesh. We extract this alignment signal and project these directions back onto the mesh surface. Using these projected directions as constraints combined with a smoothness energy, we solve a least squares system to obtain a per-face cross field that is both feature aligned and smooth.

Our method is a modular approach for extracting directional signals from 2D images and thus is compatible with different 2D image generation models (\cref{fig:diff-priors}). 
Since these models are visually informed, our final cross fields and corresponding quad meshes closely align with salient visual features (\cref{fig:teaser}). More generally, our method can be applied to any 2D image with gradient directions, such as coarse hand-drawn lines on 2D renders of the mesh (\cref{fig:user-lines}). This 2D hand-drawn guidance provides additional fine-grained control without requiring in-depth 3D modeling skills or technical knowledge. We evaluate our method on a variety of geometries ranging from organic shapes to mechanical objects and produce superior quad feature alignment over existing baseline methods.

\section{Related Work}
A large body of work aims to generate feature-aligned cross fields for downstream tasks such as quad meshing. These works can largely be divided into classical energy-based optimizations~\cite{Crane:2010:TCD, 10.1145/1531326.1531383, 10.1145/1356682.1356683}, typically prioritizing smoothness and curvature alignment~\cite{10.1145/2461912.2462005, 10.1145/1531326.1531383}, and more modern neural methods that provide continuous representations of these classical optimizations~\cite{dong2025neurcrossneuralapproachcomputing}, and/or enable data-driven feed-forward generation~\cite{dong2025crossgen, https://doi.org/10.1111/cgf.14366,
NeuFrameQ-25}.

\subsection{Classical and Optimization Methods}
The majority of existing cross-field design methods create cross-fields by optimizing mathematical energies~\cite{10.1145/3084873.3084921}. Methods such as \cite{Panozzo:2014, 10.1145/2766906, 10.1145/1356682.1356683, 10.1145/1531326.1531383, Crane:2010:TCD, 10.1145/2461912.2462005, Jakob2015Instant, 10.1111:cgf.13498} optimize energies that include smoothness of the cross field as a primary component.

Many optimization methods rely on geometric quantities as a stand-in for feature alignment. The most common geometric quantity used to approximate feature alignment is the mesh's principal curvature directions~\cite{10.1145/1531326.1531383, 10.1111:cgf.13498}. For example, Power Fields ~\cite{10.1145/2461912.2462005, azencot2017consistent} includes alignment with the mesh's principal curvature directions as one of two terms in its optimization. More recently, NeurCross  ~\cite{dong2025neurcrossneuralapproachcomputing} expands on this concept by training a neural network to predict a signed distance function representing a continuous relaxation of the mesh. NeurCross then aligns their cross field to the curvature of this continuous relaxation. Other methods, such as Octahedral Frames ~\cite{zhang2020octahedralframesfeaturealignedcrossfields} argue that different geometric quantities, such as sharp creases, provide better alignment to sharp features than principal curvature.

Despite the widespread use of these geometric quantities, feature alignment is inherently semantic, so no geometric quantity can perfectly guarantee feature alignment. Our method improves on these shortcomings by including in our optimization a direct representation of the semantic notion of feature alignment through information extracted from generative 2D priors.

As an alternative option to improve feature alignment, many prior works allow users to manually provide additional constraints as inputs to guide the optimization. Some prior works allow users to specify the locations and indices of the cross field's singularities ~\cite{Crane:2010:TCD}, particular edges for the cross field to strongly align with ~\cite{10.1145/3450626.3459941}, or other guiding constraints that are tied to specific geometric primitives ~\cite{viertel2018approachquadmeshingbased, 10.1145/1531326.1531383, 10.1145/2461912.2462005}. Techniques such as these are beneficial in allowing users fine control over the optimization, but they are inherently tedious and labor intensive, as they require selecting particular vertices, edges, and faces out of potentially thousands on the mesh's surface.

Our optimization provides automatic feature alignment without the need for users to manually specify additional inputs. Additionally, for users who want more fine control over feature alignment, our method provides a much easier experience by extracting features from sketches of grid lines on renders. 

\begin{figure}
    \centering
    \includegraphics[width=\linewidth]{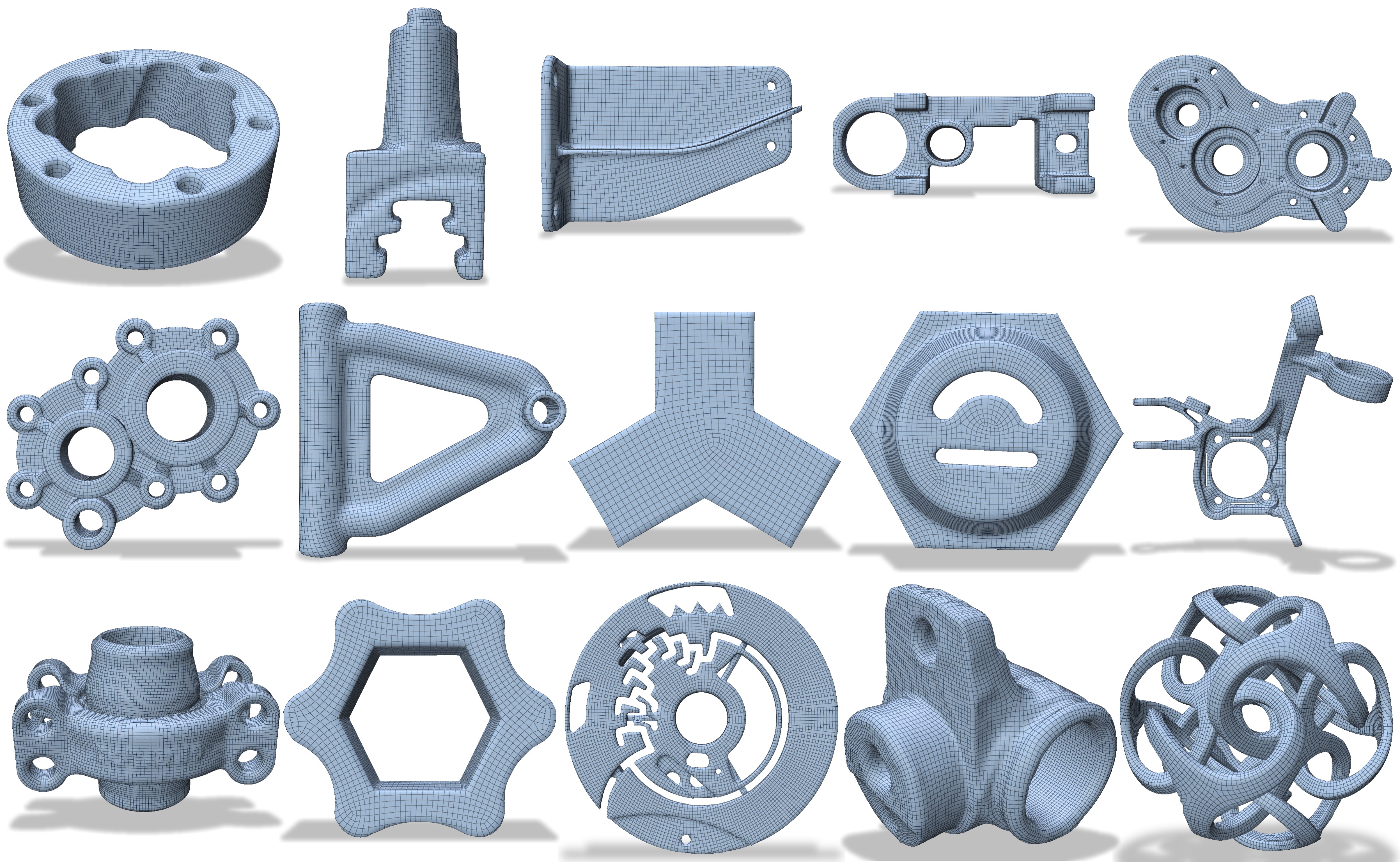}
    \caption{\textbf{Mechanical Shapes}. Our visually-guided approach produces feature-aligned results on mechanical, CAD-produced objects. These shapes are often complex and contain sharp and thin features. Our alignment is obtained without any sharp feature constraints and is purely dictated by the 2D guidance.}
    \label{fig:mechanical-shapes}
\end{figure}

\begin{figure*}[t]
    \centering
    \includegraphics[width=0.95\linewidth]{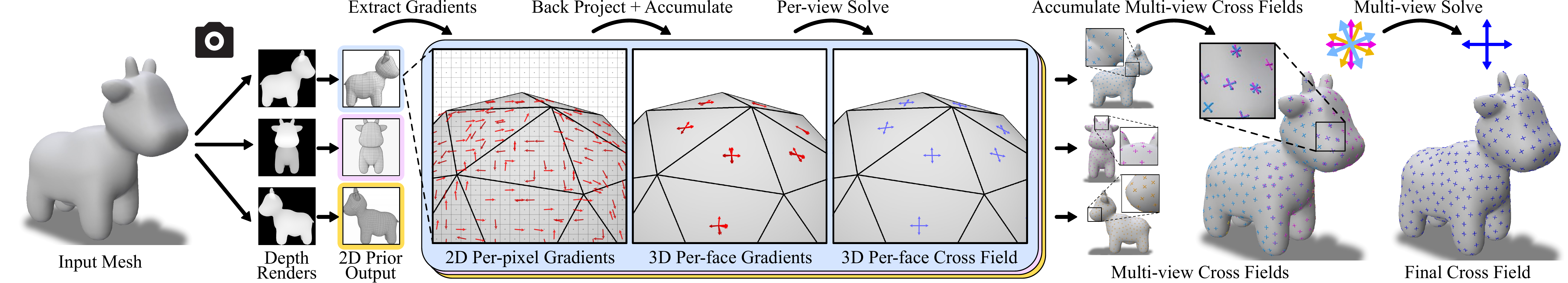}
    \vspace{-2mm}
    \caption{\textbf{Overview}.
    Taking a mesh as input, our method renders depth from multiple views using these to condition a 2D generative model which produces multi-view images depicting a quad pattern on the mesh. We extract the alignment directions from these images using pixel gradients and project those directions back to the 3D surface. Here, we interpolate using two solves, first per-view to accumulate the directions into a cross field over the faces and second to interpolate these multi-view cross fields into a single smooth cross field for the mesh.}
    \label{fig:overview}
\end{figure*}

\subsection{Data-driven Methods}
Recently, researchers have begun incorporating data-driven methods to quad meshing ~\cite{dong2025crossgen, NeuFrameQ-25, https://doi.org/10.1111/cgf.14366}, where models learn patterns from large datasets to guide the quad meshing process. These methods can enable improved feature alignment, as patterns in the input data can encode the semantic features. However, most existing data-driven methods generate quad meshes from scratch ~\cite{Li_2025, hao2024meshtronhighfidelityartistlike3d} and there are few data-driven methods for field-guided quad meshing of existing meshes.

CrossGen~\cite{dong2025crossgen} proposed a data-driven approach for field-guided quad meshing by distilling results from NeurCross into a feed-forward model. As such, CrossGen was trained directly on NeurCross data. Thus, it does not learn semantic patterns from artist created meshes and faces the same drawbacks with regards to feature alignment as NeurCross. Learning Direction Fields ~\cite{https://doi.org/10.1111/cgf.14366} describes a framework for training a model to predict cross fields for quad meshing trained on real artist-created data. While results are promising, the model is trained on a dataset consisting of only 10 human meshes in various poses and thus cannot generalize to arbitrary objects.

NeuFrameQ~\cite{NeuFrameQ-25} introduces a promising method for data-driven cross field-guided quad meshing. The authors propose a neural network that operates on point clouds and directly predicts frame fields (an abstraction of cross fields). Most importantly, NeuFrameQ trains on an extensive dataset of over $270$k artist-created meshes. Since the meshes in its training data were generated directly by artists, they contain alignment to nuanced semantic features not necessarily present in the geometry. Furthermore, the scale of the training dataset allows for far better generalization than existing methods. However, the semantic understanding of NeuFrameQ is still limited by the relative scarcity of 3D data. In contrast, our approach leverages pre-trained 2D generative models which have been trained on billions of images and have demonstrated impressive general 3D understanding~\cite{du2023generative}. Using these models as guidance, our method is able to generalize to arbitrary shapes while aligning quads to nuanced semantic features of the objects.

\section{Method}

\paragraph{Overview} \ourmethod{} produces a feature aligned face-based cross field $\bm{F}$ for an input triangle mesh $M$ using priors implicitly contained in 2D generative models. This cross field can then be used to extract a high-quality quad mesh $M'$. Our method uses a depth-conditioned text-to-image model to produce images of a feature-aligned quadrangulated mesh from multiple views. We extract the 2D quad alignment directions implicitly contained in these images and project them back onto the 3D surface. We then perform a two-stage interpolation consisting of linear solves using candidate alignment directions as constraints alongside a smoothness energy. Our first solve aggregates the projected directions to a per-face cross field for each view, and our second solve consolidates these multi-view cross fields into a single field $\bm{F}$ that is both smooth and feature-aligned. Finally, using this cross-field $\bm{F}$, we extract a quad mesh $M'$. See \cref{fig:overview} for an overview of our method.

\subsection{Generating Feature Aligned 2D Quad Mesh Images}
\label{subsec:generation}
\begin{figure}[b]
    \centering
    \includegraphics[width=\linewidth]{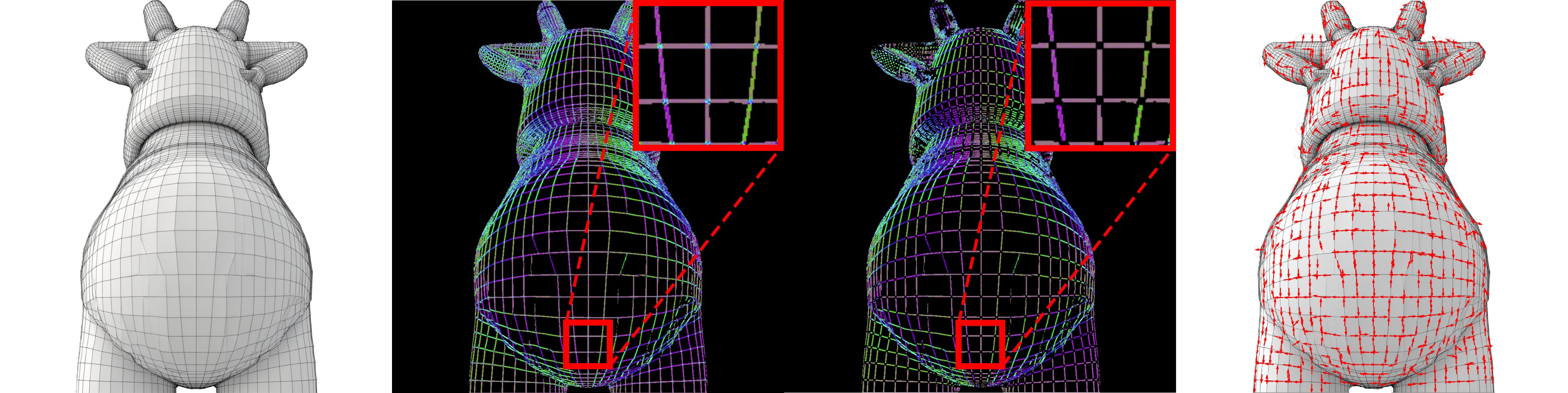}
    \caption{\textbf{Gradient Extraction}.
    To extract alignment directions from our 2D guidance images (left), we first apply a Scharr kernel~\cite{Scharr2000OptimaleOI} to get per-pixel gradients.
    We threshold on magnitude to filter for only gradients on mesh grid lines (middle left).
    We encode gradient directions with colors in a 4-rotationally symmetric color scheme.
    As grid lines are continuous, nearby pixels should have similar colors.
    We remove erroneously averaged and non-smooth gradients at grid-line intersections by filtering for coherence (middle right), leaving us with 
    directions corresponding to the edge flow from the guidance image (right).}
    \label{fig:grad-extraction}
\end{figure}

We build our feature alignment directly from 2D generative models, which create a quadrangulated wireframe of our mesh from several prescribed views. To do so, we select $B=6$ canonical views (left, right, front, back, top, bottom) and render depth maps of the mesh from these views to get $D \in \mathbb{R}^{B \times H \times W \times 1}$. We also define a prompt $y$ that describes our quad mesh texture (see \cref{sec:implementation-details} for specific prompt details). We then use a ControlNet~\cite{zhang2023adding} model conditioned on both the prompt $y$ and the depth maps $D$ to generate images $I \in \mathbb{R}^{B \times H \times W \times 3}$ that depict a wireframe quad mesh texture on our mesh from each view.
Once we have obtained the generated images $I$, we can use these as guidance for our cross field by aligning our field to the quads depicted in the images (see \cref{fig:grad-extraction,fig:all-steps}).

\subsection{Image Gradient Extraction}
\label{subsec:grad-extract}
To determine the alignment of the quads in each generated image, we compute color gradients at each pixel. We apply a Scharr kernel \cite{Scharr2000OptimaleOI} (in both the $u$ and $v$ directions): 
\[
    k_u = \begin{bmatrix}
            -3 & 0 & 3\\
            -10 & 0 & 10\\
            -3 & 0 & 3\\
          \end{bmatrix}, \quad k_v = k_u^T
\]
over all pixels in all images in the batch. This gives us per-pixel partial derivatives of the image with respect to $u$ and $v$,
\[
   G_u = k_u * x_0, \quad G_v = k_v * x_0 \quad \in \mathbb{R}^{B \times H \times W}
\]
respectively. We encode these as a single complex-valued array $G = G_u + iG_v \in \mathbb{C}^{B \times H \times W}$, where each entry represents the gradient vector at the corresponding pixel as a complex number with the $u$ component as its real part and the $v$ component as its imaginary part. We also rotate $G$ by $90^\circ$ so that our vectors are aligned with (and not perpendicular to) the grid lines in the image. This rotation is necessary since we cannot treat the image-space gradients as $4$-\textit{RoSy} due to projection (see \cref{fig:solve-ablation,subsec:ablations}).

The extracted gradients $G$ contain our alignment signal at pixels near the grid lines of the quads depicted in $I$. However, at all other pixels, the directions contained in $G$ are not meaningful. Since meaningful gradients coincide with the sharp color changes between the dark grid lines and the light mesh surface, we can isolate them by thresholding on $|G|$. Specifically, we discard gradients with magnitude below $12\%$ of the per-image maximum. This removes most of the gradients on pixels that do not correspond to the quad grid lines. However, at quad corners, some pixel gradients still have high magnitude, but are not aligned with the quads. These problematic gradients appear because the direction of maximal color change at grid line intersections is the average direction of the two orthogonal edges. To address this, we also threshold the gradients using a measure of how aligned the gradients are to their neighbors. Specifically, we first compute the smoothed structure tensor
\[
    S = \begin{bmatrix}
            \langle G_u^2 \rangle & \langle G_u G_v\rangle\\
            \langle G_u G_v \rangle & \langle G_v^2 \rangle\\
        \end{bmatrix}
\]
where $\langle \cdot \rangle$ represents a Gaussian blur with an $11 \times 11$ kernel. Letting $\lambda_1 > \lambda_2$ be the eigenvalues of $S$, we define the per-pixel coherence as:
\begin{equation}
    \mathrm{coh} = \frac{\lambda_1 - \lambda_2}{\lambda_1 + \lambda_2}.
\end{equation}
We then threshold our gradients a second time, discarding gradients with coherence below $0.5$. An overview of the gradient extraction process is shown in \cref{fig:grad-extraction}.

\subsection{Gradient Back Projection}
To inform our 3D cross field, we back project our extracted image-space gradient vectors to the surface tangent space. Let $M = (V, F, E)$ consist of vertices $V \in \mathbb{R}^{n\times3}$, faces $F \in \mathbb{N}^{m\times3}$ and edges $E \in \mathbb{N}^{l\times2}$. We associate with each face $f \in F$ with a pair of orthogonal basis vectors $B_f = \begin{bmatrix}B_{f, x}&  B_{f, y}\end{bmatrix}$ to describe points in its 2D tangent plane. Now, consider any pixel in the input image and its corresponding world space position $p = (x, y, z)$ on face $f \in F$ of the mesh. We construct the Jacobian of the camera projection at $p$ as
\begin{equation}
    J_p = \begin{bmatrix}
            \frac{f_x}{z} & 0 & \frac{-f_x x}{z^2}\\
            0 & \frac{f_y}{z} & \frac{-f_y y}{z^2}\\
        \end{bmatrix}
\end{equation}
where $f_x$ and $f_y$ are the camera intrinsic properties focal width and focal length respectively.
Let $G_{\text{tan}}$ and $G_{\text{img}}$ represent the surface tangent space and image space gradients respectively for $p$. Then, since $G_{\text{img}}$ represents the projection of the surface gradients $G_{\text{tan}}$, we can write it as a function of $G_{\text{tan}}$ and the Jacobian of our camera projection:
\begin{equation}
    J_pM_{\text{WtoC}}B_{f}G_{\text{tan}} = G_{\text{img}}
\end{equation}
where $M_{\text{WtoC}}$ is the world to camera space transformation matrix. Finally, to obtain our tangent space gradients, we solve for $G_{\text{tan}}$:
\begin{equation}
    G_{\text{tan}} = \left(J_pM_{\text{WtoC}}B_{f}\right)^{-1}G_{\text{img}}.
    \label{eq:unproj}
\end{equation}
Using this process, we can then back project all image-space gradients by rasterizing the mesh from each view and applying \cref{eq:unproj} in parallel to each pixel gradient. From this we obtain surface tangent-space gradients $G_{\text{tan}} \in \mathbb{C}^{B \times m \times k}$ where $k$ is the maximum number of gradients stored on any face. Since faces contain a variable number of gradients, we pad all faces containing fewer than $k$ gradients with $0$s.

\subsection{Smooth Interpolation}
\label{subsec:smooth-interp}
\begin{figure}
    \centering
    \newcommand{\pl}{-4}
    \newcommand{\two}{-8}
    \begin{overpic}[width=\linewidth]{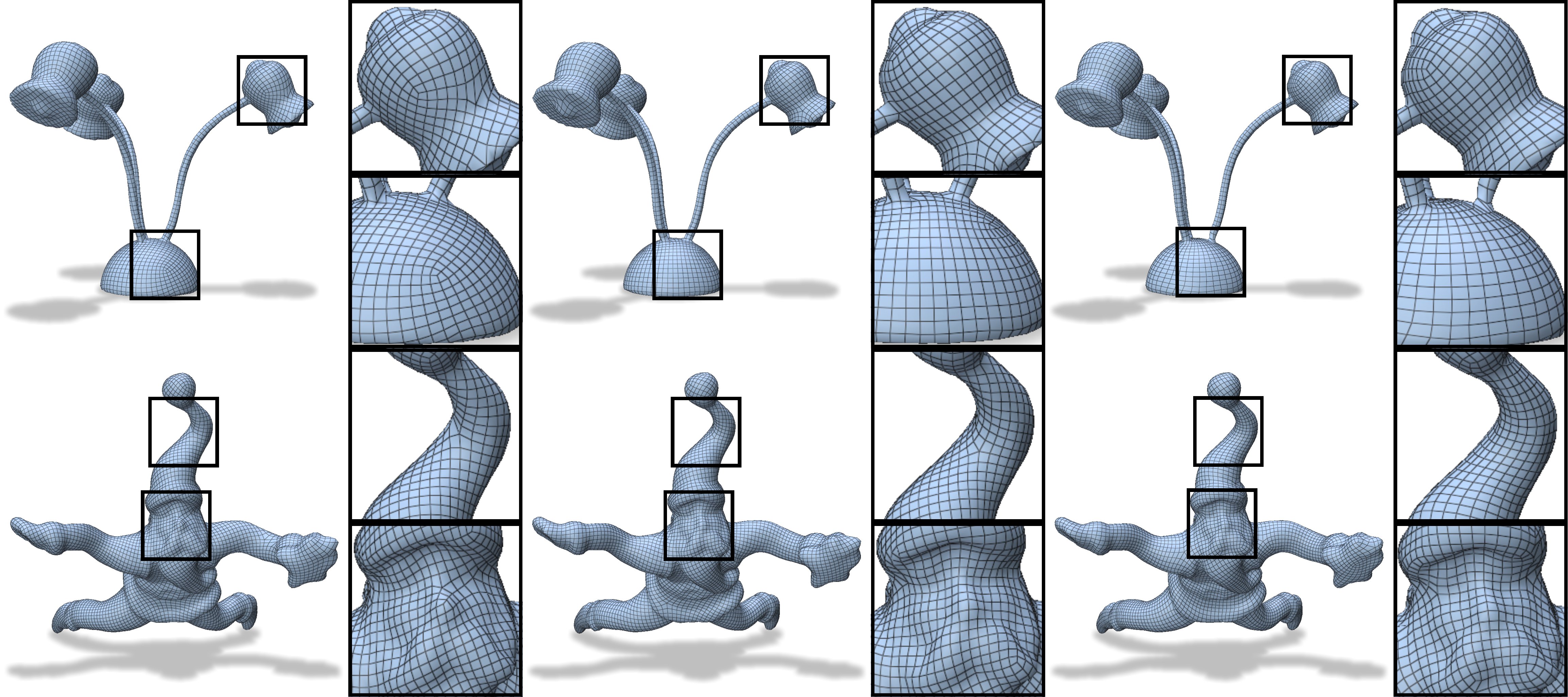}
    \put(5,  \pl){\textcolor{black}{Multi-view Solve}}
    \put(36,  \pl){\textcolor{black}{Image Interpolation}}
    \put(36,  \two){\textcolor{black}{+ Multi-view Solve}}
    \put(69,  \pl){\textcolor{black}{\textbf{Ours} (Surface Solve}}
    \put(69,  \two){\textcolor{black}{+ Multi-view Solve)}}
    \end{overpic}
    \vspace{1mm}
    \caption{\textbf{Interpolation Ablation.} Back-projecting all gradients and performing a single surface interpolation (left) can cause views with denser gradient extraction to dominate the final solve. A two-stage interpolation that first interpolates $4$-\textit{RoSy} quantities in image space introduces projection artifacts, since orthogonal surface vectors are not guaranteed to remain orthogonal in the image. Our two-stage approach back-projects gradients and interpolates them directly on the surface, yielding dense per-view signals for the multi-view solve. This prevents projection artifacts and ensures equal contribution from all views.}
    \label{fig:solve-ablation}
\end{figure}
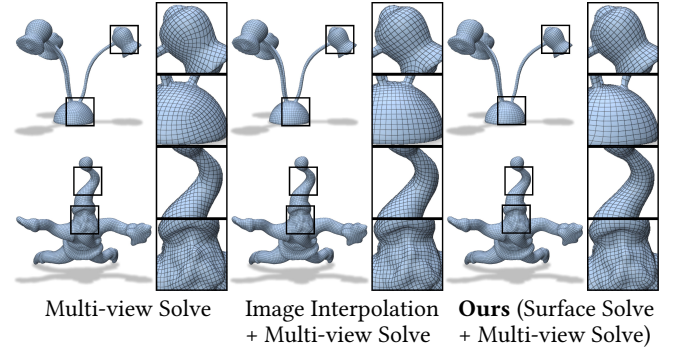

We interpolate our back-projected surface tangent-space gradients $G_{\text{tan}}$ into a per-face cross field $\bm{F}$. We perform this interpolation in $4$-\textit{RoSy} space to account for the $90^\circ$ rotational symmetry of crosses. Given a field of alignment vectors $\bm{u}$, we represent its $4$-\textit{RoSy} cross field with $\bm{F} = \bm{u}^4$. This \textit{power field} representation~\cite{10.1145/2461912.2462005,azencot2017consistent} identifies $\bm{u}$ with its three $90^\circ$ rotations. Importantly, we perform this $4$-\textit{RoSy} conversion after back projecting gradients to the surface since orthogonal vectors on the surface are not necessarily orthogonal in image space (see \cref{fig:solve-ablation}).

After back projection, some faces visible from a given view may contain multiple valid gradients, while others may contain none due to the filtering in \cref{subsec:grad-extract}. This problem compounds in the multi-view setting as faces visible across multiple views can accumulate gradients from each of those views, while faces that are always occluded contain no signal (see \cref{fig:smoothness-term}). To consolidate these sparse and competing alignment signals into to a single unified cross field, we employ the ``globally-optimal'' paradigm~\cite{10.1145/2461912.2462005}, which smoothly interpolates crosses between faces by solving a constrained linear system. We apply this solve twice: first, to unify the alignment signal within each view individually, and second, to consolidate the alignment across all views into single, consistent cross field. Our first solve processes each view $b \in B$ individually, taking the surface gradients for that view $G_{\text{tan}}^b \in \mathbb{C}^{m \times k}$ as input and producing a per-face cross field $\bm{F}^b$ for the given view. Our second solve combines the $\bm{F}^b$ fields from each view into our final cross field $\bm{F}$.

Since the linear systems in both interpolation steps share the same structure and differ only in their alignment constraints and weights, we first describe the solve for arbitrary constraint and weight values. For a system with $C$ constraints, each constraint $c \in C$ consists of a target cross alignment $\bm{F}^*_c$, a weight $w_c > 0$, and an associated face $f_c$. Some faces may have multiple constraints in $C$, while other may have none. The solve employs two separate energy terms to obtain the interpolated per-face cross field $\bm{F} \in \mathbb{C}^{m}$. First, the smoothness energy ensures that crosses are aligned with their neighbors:
\begin{equation}
    E_s(\bm{F}) = \sum_{e=(f,g) \in E} w_e \left|\bm{F}_f \bar{e}\,^4_f - \bm{F}_g \bar{e}\,^4_g \right|^2.
\end{equation}
In this energy, $w_e$ are harmonic weights given by \cite{brandt2018modeling}, and $\bar{e}_f$ and $\bar{e}_g$ are the complex conjugates of the representations of the vector along edge $e$ as a unit-length complex number relative to the bases of faces $f$ and $g$ respectively. These edge terms serve to transport $\bm{F}_f$ and $\bm{F}_g$ to the same space so comparisons are valid. 
Second, the alignment energy encourages crosses in the solution to align with the provided constraints:
\begin{equation}
    E_c(\bm{F}) = \sum_{c \in C} w_c \left|\bm{F}_{f_c} - \bm{F}^*_c\right|^2.
\end{equation}
The final solve aims to find
\begin{equation}
    \bm{F} = \text{argmin} \left[
        \lambda_s E_s(\bm{F})
        +
        \lambda_c E_c(\bm{F})\right]
    \label{eq:powerfields}
\end{equation}
where $\lambda_s$ and $\lambda_c$ are user-defined parameters that control the ratio between smoothness and alignment. The result of this solve is a cross field $\bm{F}$, which we then normalize in accordance with \cite{10.1145/2461912.2462005} by setting $\bm{F}_f = \bm{F}_f / | \bm{F}_f|$ to get our final unit-length crosses for each face. We discuss the specific constraints and weights used for the first (per-view) solve in ~\cref{subsec:per-view-crosses} and second (multi-view) solve in \cref{subsec:multi-view-crosses}.

\subsection{Per-View Interpolation}
\label{subsec:per-view-crosses}
To convert our back-projected surface gradients into a smooth field of crosses defined on faces, we apply the solve defined in \cref{subsec:smooth-interp} to each view individually. We use the $4$-\textit{RoSy} ``power fields'' representation of the surface gradients as our constraints, setting $\bm{F}^{*} = (G_{\text{tan}}^{b})^4$. For a given constraint $\bm{F}_c$ that corresponds to a surface gradient at point $p$ on face $f$, we assign it face $f$ and set the constraint weight to
\begin{equation}
    w_c = \exp\left(\frac{-d_p^2}{2 \sigma_{f}^2}\right)
    \label{eq:wc1}
\end{equation}
where $d_p$ is the distance between $p$ and $f$'s centroid and $\sigma_f$ is the distance of the farthest vertex from the centroid. These weights use a standard Gaussian kernel to smoothly decay with distance from the centroid, and $\sigma_f$ ensures results are normalized with respect to the triangle's size. Additionally, when applying this solve to a given view, we ensure that non-visible faces cannot have any affect on the interpolation by excluding edges adjacent to non-visible faces from the smoothness term in the solve.

Solving \cref{eq:powerfields} with these constraints and weights, we obtain a per-face cross field $\bm{F}^b$ for each view. Interpolating the sparse surface gradients to mesh faces for each view ensures a uniform density alignment signal across all views. Thus, each view contributes equally to the multi-view solve in \cref{subsec:multi-view-crosses}, regardless of its original image-space gradient density (see \cref{fig:solve-ablation}).

\subsection{Multi-View Interpolation}
\label{subsec:multi-view-crosses}
Using the per-view fields obtained in \cref{subsec:per-view-crosses}, we perform a second solve to obtain a single, consistent cross field for the mesh. We use the concatenation of all $B$ of our $\bm{F}^b$ cross fields as our constraints, setting $\bm{F}^* = \big\Vert_{b \in B} \bm{F}^b$. Each $f_c$ is assigned as the face on which the given cross is defined (due to the concatenation, $f_c$ is not unique). For this solve, we also assign a view $b_c$ for each constraint $c$ as the view from which the constraint $c$ was sourced. The weight for each constraint is determined by combining two different proxies for our confidence in its alignment.

First, we prioritize crosses on faces that are aligned with the original view direction that the cross was extracted from. We compute these weights as
\begin{equation}
     w^{\text{view}}_c = \left ( \ell_{b_c} \cdot N_{f_c} \right )^2
\end{equation}
where $\ell_{b_c}$ is the look at vector pointing towards the camera for the view $b_c$, and $N_{f_c}$ is the normal vector of face $f_c$. When faces are not aligned with the view, the grid lines are more distorted by perspective and more difficult to see in the image, so the extracted gradients and their corresponding crosses are less trustworthy. These weights compensate for such crosses by introducing additional smoothness to correct any potential noise caused by this lack of visibility.

Second, we add a weight term to reconcile conflicting alignment between views. Let $C_{f} = \{c \in C : f_c = f\}$ denote the set of constraints on face $f$. For each face $f$, we define the multi-view coherence weight as
\begin{equation}
    w^{\text{coh}}_f = \frac{\left| \sum_{c \in C_f} w^{\text{view}}_c \, \bm{F}^*_c \right|}{\sum_{c \in C_f} w^{\text{view}}_c}
\end{equation}
and assign each constraint the coherence weight of its associated face:
\begin{equation}
    w^{\text{coh}}_c = w^{\text{coh}}_{f_c}.
\end{equation}
These weights are near $1$ for constraints on faces in which the crosses are similar across all views, and near $0$ for constraints on faces that are seen by multiple views with conflicting crosses. In cases where multiple views have conflicting crosses on a given face, this weight serves to decrease the strength of alignment term, thus allowing the smoothness term to dominate and resolve the ambiguity.

To get our final constraint weight, we take the product of the two weight terms:
\begin{equation}
    w_c = w^{\text{coh}}_c w^{\text{view}}_{c}.
\end{equation}
A visualization of this multi-view interpolation is shown in \cref{fig:smoothness}.

\begin{figure}
    \centering
    \newcommand{\pl}{-4}
    \newcommand{\snd}{-8}
    \begin{overpic}[width=\linewidth]{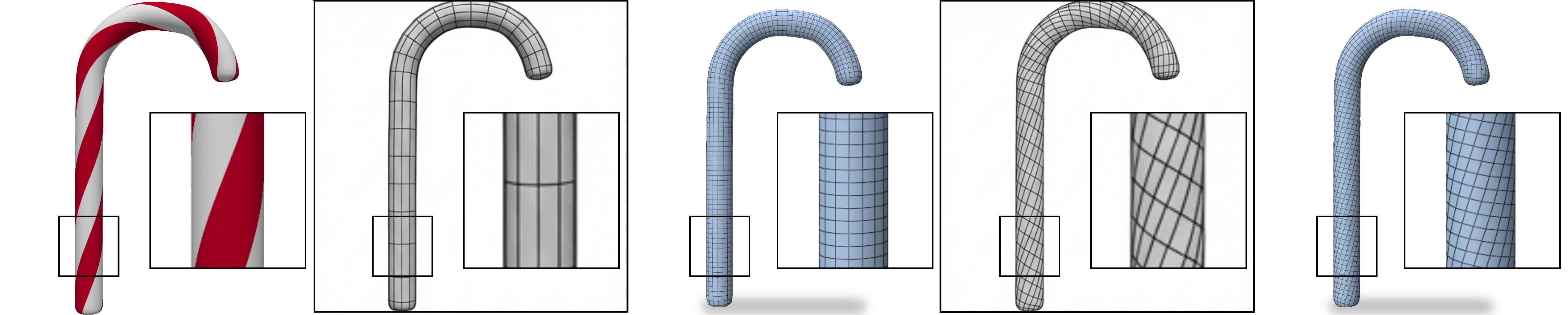}
        \put(5,  \pl){\textcolor{black}{Texture}}
        \put(31,  \pl){\textcolor{black}{w/o Texture}}
        \put(73,  \pl){w/ Texture}
    \end{overpic}
    \vspace{-0.2cm}
    \caption{\textbf{Texture-Aligned Quads}. Given a textured mesh (left), we show the 2D prior output (gray) and final quad mesh (blue) using either untextured (middle) or textured (right) renders of the mesh to condition our 2D prior. Using untextured renders produces quads aligned to the features of the shape, while using textured renders allow our method to align quads to the features in the texture.}
    \label{fig:texture-alignment}
\end{figure}

\section{Experiments}

We demonstrate the capabilities of \ourmethod{}, applying it to a diverse set of meshes (\cite{turbosquid, coseg_2011, 10.1145/3450626.3459941}) consisting of both organic and mechanical shapes. \cref{subsec:extraction} gives additional details on how we convert our cross fields into quad meshes. In \cref{subsec:visual-alignment}, we show how our visual guidance enables strong semantic feature alignment in our quad meshes. In \cref{subsec:eval}, we evaluate our approach against existing quad meshing techniques, showing improved performance over baseline methods both qualitatively and quantitatively. \cref{subsec:ablations} justifies key method components through an ablation. We discuss applications of our approach in \cref{subsec:applications}. Finally, we highlight key properties and the robustness of our method in \cref{subsec:robustness}. In our experiments, we use the Flux~\cite{labs2025flux1kontextflowmatching} model as our 2D image prior.

\subsection{Quad Mesh Extraction}
\label{subsec:extraction}
To produce a quad mesh from a per-face cross field, we integrate the cross field to obtain a parameterization and then extract a quad mesh by placing quad vertices and edges on the parameterization integer grid and iso-lines respectively. To integrate our field, we first run iterations of Directional's~\cite{Directional} polyvector curl correction with the algorithm described in \cite{meekes2021unconventional} to ensure that our field is integrable, while still retaining alignment to our original field. We pass this integrable field to libigl~\cite{libigl} to perform seamless integration using MiQ~\cite{bommes2009mixed}. This output parameterization is then passed to libQEx~\cite{Ebke:2013:QRQ:2508363.2508372} to extract the final quad mesh. While the integration usually completes in under $5$ minutes, it is sensitive to the mesh quality and can take a prohibitively long time to produce a parameterization on fields with many singularities. In severe cases where MiQ is unable to produce a parameterization within $4$ hours, we fall back to using QuadWild~\cite{10.1145/3450626.3459941} to extract a quad mesh conditioned on the given cross field. These settings are used for all methods that produce a cross fields as the final output (Ours, NeurCross).

\begin{figure*}[t]
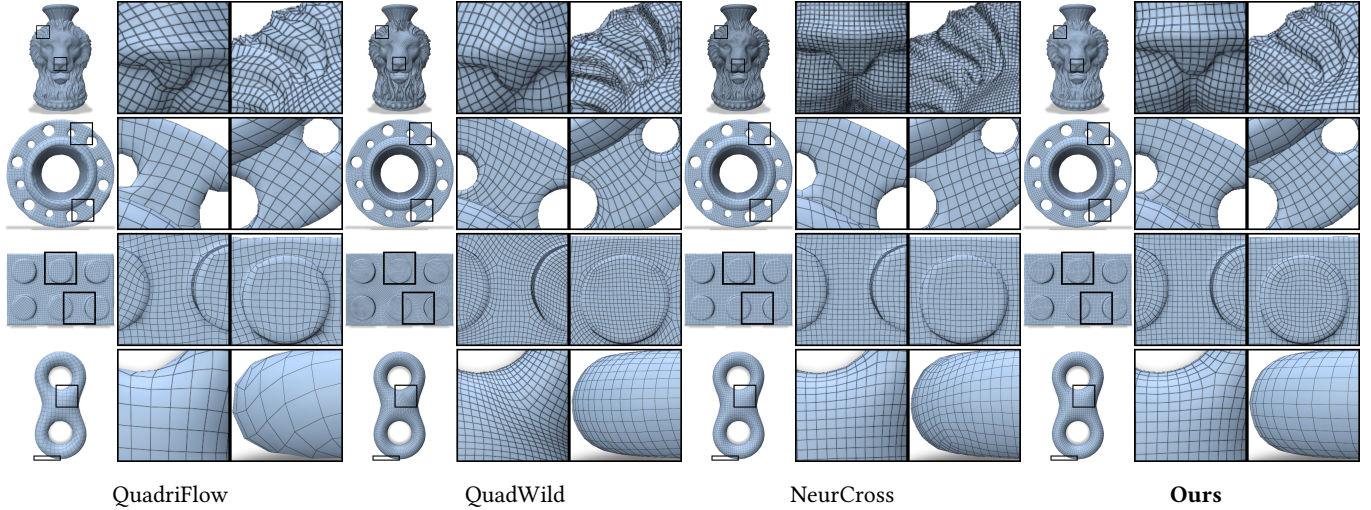

    \centering
    \newcommand{\pl}{-3}
    \begin{overpic}[width=\linewidth]{figures/images_compressed/qual_comp.jpg}
    \put(8,  \pl){\textcolor{black}{QuadriFlow}}
    \put(34,  \pl){\textcolor{black}{QuadWild}}
    \put(58,  \pl){\textcolor{black}{NeurCross}}
    \put(86,  \pl){\textcolor{black}{\textbf{Ours}}}
    \end{overpic}
    \vspace{0 mm}
    \caption{\textbf{Qualitative Comparison}. We compare our method to QuadriFlow~\cite{10.1111:cgf.13498}, QuadWild~\cite{10.1145/3450626.3459941} and NeurCross~\cite{dong2025neurcrossneuralapproachcomputing}. QuadriFlow and QuadWild often produce wavy edge flow, while NeurCross tends to produce overly axis-aligned edge flow that fails to align to specific features. In contrast, our method produces edge flow that demonstrates superior alignment to both semantic and geometric features.}
    \label{fig:qual-comp}
\end{figure*}

\subsection{Importance of Visual Guidance}
\label{subsec:visual-alignment}

\noindent\textbf{Visual Feature Alignment.}\quad
The 2D models we use to guide our alignment have strong semantic and geometric understanding. This property enables the alignment signal we extract to capture both local geometric features of the shape such as sharp edges (see \cref{fig:mechanical-shapes}) and curvature as well as more global semantic features such as axes of symmetry. In \cref{fig:teaser}, we show how the quad meshes we generate capture both of these components. In cases where the geometric signal is clear such as the human statue (left), our method produces grid lines that flow along the directions of principal curvature. Conversely, on the flat surface of the Lego brick (right), the principal curvature is not well defined. At locations on this flat surface in between the raised studs, the nearby geometric features on the edges of the studs exhibit radial symmetry. However, our method still produces grid lines on this flat surface of the brick that align to primary axes of the shape (up/down and left/right for this rectangle), indicating global semantic understanding that cannot be obtained from the local geometric features used to guide existing methods (see \cref{fig:qual-comp}, row $3$).

\smallskip
\noindent\textbf{Generality.}\quad
Since our method is guided by 2D image models, we are able to extract alignments for arbitrary objects in the distribution of the model. Thus, our approach is not restricted to any particular class or category of input mesh. In \cref{fig:gallery,fig:mechanical-shapes}, we show the results of our method applied to a diverse collection of input meshes, spanning both organic shapes with smoother curves and less prominent geometric features as well complex mechanical shapes with large flat surfaces, sharp edges, and thin features. In both cases, our method produces desirable alignment dictated by the 2D guidance, without any explicit geometric dependence.

\subsection{Evaluation}
\label{subsec:eval}

We compare our method both qualitatively and quantitatively to QuadriFlow~\cite{10.1111:cgf.13498}, QuadWild~\cite{10.1145/3450626.3459941}, and NeurCross~\cite{dong2025neurcrossneuralapproachcomputing} on the QuadWild300 dataset.

\smallskip
\noindent\textbf{Qualitative Comparison.}\quad
We show in \cref{fig:qual-comp} that our quads exhibit better alignment to both semantic and geometric features than baselines methods. Method such as Quadriflow and QuadWild tend to produce meshes with wavey edge flow while NeurCross produces overly rigid grids that fail to align to specfic features. For example, in shapes with clear semantic features such as the lion vase (row $1$), our method (right) gives edges that flow along and across the nose, obeying the axes of symmetry of the shape and matching up with the nose tip. In contrast, other methods' edges flow diagonally over the nose (QuadriFlow) and introduce singularities in the mesh grid (QuadWild). While NeurCross does align the quads to the nose, it orients nearly all quads to the vertical and horizontal axes, failing to align quads to the features of the mane as ours does. On shapes with less obvious semantic features such as the Lego brick (row $3$), our visual approach still results in superior feature alignment. Our method (right) produces axis-aligned edge flow on the flat brick surface, while still capturing the radially symmetric curved geometry of the raised studs. QuadriFlow (left) and QuadWild (middle left) produce wavy edge flow on the brick surface, while NeurCross (middle right) makes all quads axis-aligned and fails to match the radial symmetry on the raised studs.

\begin{figure}
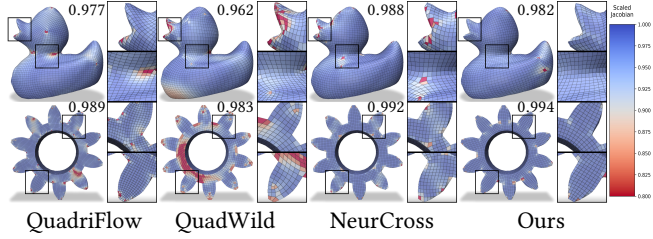

    \centering
    \newcommand{\pl}{-4}
    \newcommand{\two}{29}
    \newcommand{\three}{14}
    \begin{overpic}[width=\linewidth]{figures/images_compressed/quad_quality.jpg}
    \put(3,  \pl){\textcolor{black}{QuadriFlow}}
    \put(26,  \pl){\textcolor{black}{QuadWild}}
    \put(50,  \pl){\textcolor{black}{NeurCross}}
    \put(79,  \pl){\textcolor{black}{Ours}}
    \put(9.5,  \two){\textcolor{black}{\scriptsize$0.977$}}
    \put(32.5,  \two){\textcolor{black}{\scriptsize$0.962$}}
    \put(55.8,  \two){\textcolor{black}{\scriptsize$0.988$}}
    \put(79,  \two){\textcolor{black}{\scriptsize$0.982$}}
    \put(9.5,  \three){\textcolor{black}{\scriptsize$0.989$}}
    \put(32.5,  \three){\textcolor{black}{\scriptsize$0.983$}}
    \put(55.8,  \three){\textcolor{black}{\scriptsize$0.992$}}
    \put(79,  \three){\textcolor{black}{\scriptsize$0.994$}}
    \end{overpic}
    \vspace{-2mm}
    \caption{\textbf{Scaled Jacobian (SJ) Visualization.} We visualize the SJ of the quad mesh outputs
    QuadriFlow~\cite{10.1111:cgf.13498}, QuadWild~\cite{10.1145/3450626.3459941}, NeurCross~\cite{dong2025neurcrossneuralapproachcomputing}, and our method.
    Higher SJs (blue) indicate higher quality quads, while lower SJs (red) indicate stretching and twisting.
    We also report the mean SJ for each mesh.
    Our method (right) produces higher or comparable quality quads as baseline methods.}
    \label{fig:quad-quality}
\end{figure}

We also visualize the scaled Jacobian (SJ), a measure of quad quality in \cref{fig:quad-quality}. This value ranges from $-1$ to $1$ where a value of $1$ indicates a perfect quad, lower values reflect stretches and twists, and negative values indicate flipped triangles. For each face we compute the minimum SJ of its associated vertices, clamp them to the range $[0.8,1]$ and visualize these per-face values using the `coolwarm' color map. For the duck, other methods stretch or twist quads around the neck or beak while ours (right) retains quad quality in those regions. On the gear, our method (right) handles the radial symmetry of the shape with minimal distortion while other methods severely stretch quads in those regions. We also report the mean SJ over the entire mesh. Our method gives higher mean SJs than QuadriFlow or QuadWild indicating higher quality quads. Our mean SJs are comparable to those of NeurCross, however, looking at the figure, we can see that NeurCross's most distorted quads are visually worse than ours.

\smallskip
\noindent\textbf{Quantitative Comparison.}\quad
To evaluate the quality of the quads generated by each method, we compute the same scaled Jacobian metric over all meshes in the QuadWild dataset and report the mean value for each method (see \cref{tab:quad-metrics}). Our method produces quads with higher SJ indicating that our meshes have less stretching and twisting and are higher quality. We also report the mean percentage of irregular vertices (vertices with valence $\neq$ 4) over the QuadWild dataset to measure the smoothness of the edge flow. Our method produces the lowest mean irregular vertex percentage indicating smoother edge flow with fewer singularities.

\begin{table}[h]
\centering
\begin{tabular}{@{ }lcccc@{ }}
\toprule
Method & QuadriFlow & QuadWild & NeurCross & Ours\\
\midrule
SJ $\uparrow$ & 0.9674 & 0.9351 & 0.9737 & \textbf{0.9797}\\
IV$\% \downarrow$ & 1.4779 & 1.4807 & 1.6470 & \textbf{1.4133}\\
\bottomrule
\end{tabular}
\caption{\textbf{Quad Quality Metrics.} We compare methods on quad quality metrics and report the mean scores across the QuadWild300 dataset~\cite{10.1145/3450626.3459941}. Our method (ours) reports higher scaled Jacobian (SJ) values and lower irregular vertex (IV) percentages, indicating superior quad quality.}
\label{tab:quad-metrics}
\end{table}

\subsection{Importance of Interpolation Components}
\label{subsec:ablations}
\noindent\textbf{Interpolation Ablation.}\quad
In \cref{fig:solve-ablation}, we compare different techniques for accumulating the 2D gradient directions extracted from the image onto the surface and interpolating them into a per-face cross field. In the ``Multi-view solve'' condition (left), we back project all gradients to the surface from all views and use these directions as constraints in a solve as described in \cref{subsec:smooth-interp} to interpolate a smooth cross field. Using this approach can cause specific views to dominate the solve. If the generated image from a specific view portrays a denser quad resolution, there will be significantly more constraints in the solve from this view, which can mask the critical signals from other sparser views. On the lamp this manifests as additional singularities and diagonal grid lines, and on the Santa mesh this produces generally mis-aligned grid lines that do not adhere to features of the shape.

In the second condition (middle), we first interpolate the extracted directions using a $4$-\textit{RoSy} representation in the image space. This succeeds in densifying the image gradients so that each view contributes equally. However, $4$-\textit{RoSy} interpolation in image space is fundamentally incorrect since orthogonal vectors on the surface are not guaranteed to remain orthogonal once projected to the image coordinate system. While not as drastic, this again results in additional singularities and non-feature aligned quads.

Our two-stage approach (right) solves the challenge of equal view influence by first interpolating the back projected image gradients independently for each view to obtain dense per-view cross fields. Since this interpolation happens on the mesh surface, orthogonality is preserved. Our method then performs a multi-view solve to interpolate the per-view cross fields into a single consistent cross field for the entire mesh. This results in smooth and feature aligned quad meshes.

\begin{figure}
    \centering
    \includegraphics[width=\linewidth]{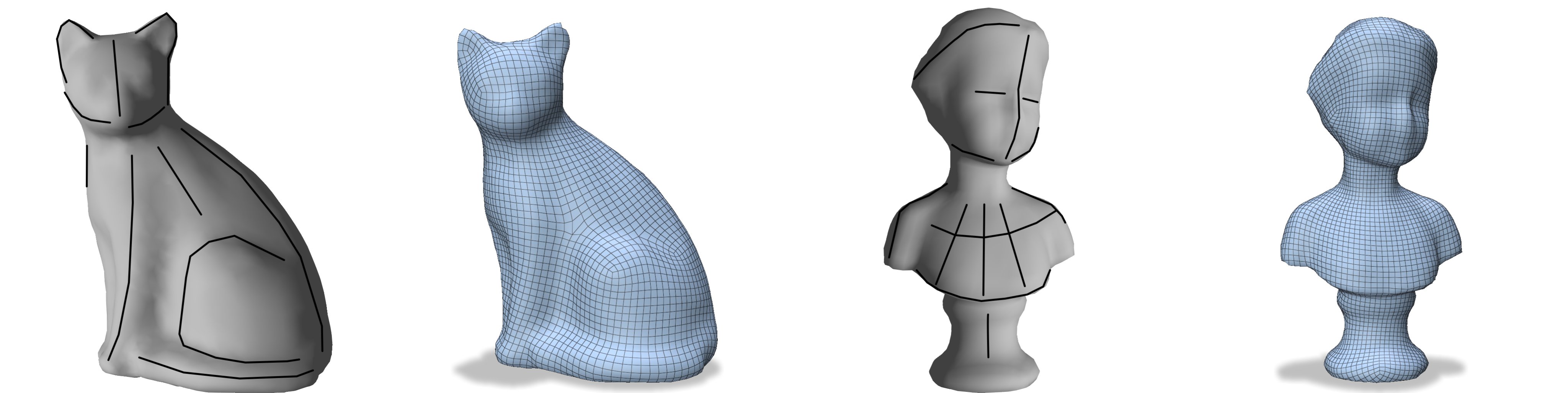}
    \vspace{-0.5cm}
    \caption{\textbf{Hand-Drawn Coarse 2D Guidance}. Our method can extract signals from coarse hand drawn alignment directions on renders of our mesh. This 2D guidance is easy to create and provides fine-grained control to the user without requiring any 3D technical knowledge or skills.}
    \label{fig:user-lines}
\end{figure}

\subsection{Applications}
\label{subsec:applications}
\noindent\textbf{Texture-Aligned Quads.}\quad
Since our \ourmethod{} method is visually informed, we can use an existing texture for the input mesh as signal to align our quads. In \cref{fig:texture-alignment}, we show a textured mesh (left) followed by our method conditioned on the untextured mesh renders (middle) and then the textured mesh renders (right). Without the texture, our method produces a quad mesh that nicely aligns to the geometry of the object. However, when we condition on the textured render of the mesh, our method is now able to align to semantics of the texture (\ie the spiraling striped barber pole pattern) even though this signal is entirely absent in the geometry. This highlights a key advantage of our method over other existing quad meshing techniques: \textit{our image-guided approach can incorporate alignment signals that exist visually, even if they have no geometric component}.

\smallskip
\noindent\textbf{User-Draw Line Guidance.}\quad
Since our method can be guided using any 2D images with alignment signal, we show an application to guidance with user-drawn lines (see \cref{fig:user-lines}). These lines can be coarse and do not require 3D skills or experience to produce. From this sparse signal, our method is able to generate a quad mesh with edge flow that closely matches the user lines such as the leg of the cat or the forehead of the bust.

\begin{figure}
    \centering
    \newcommand{\pl}{-5}
    \begin{overpic}[width=\linewidth]{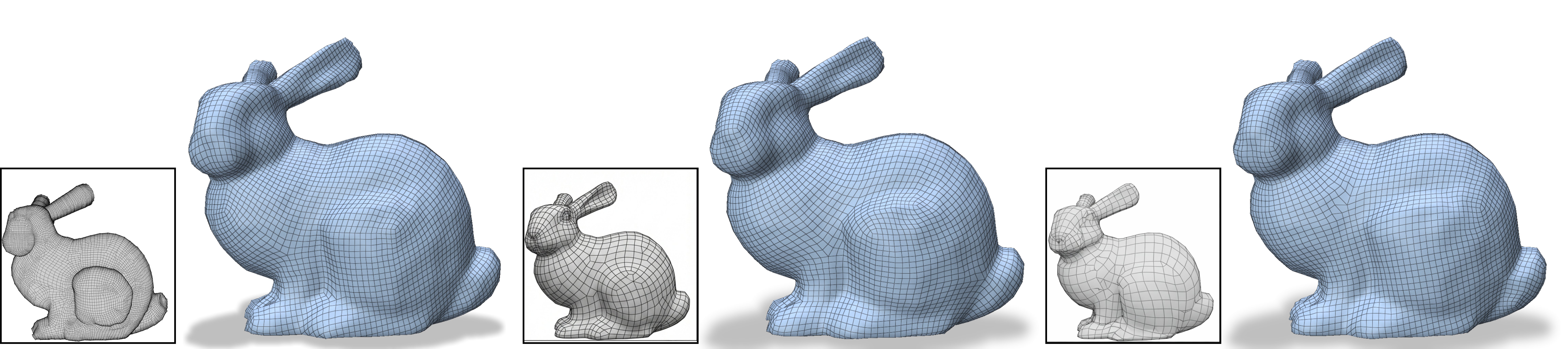}
    \put(14,  \pl){\textcolor{black}{Flux}}
    \put(43,  \pl){\textcolor{black}{Gemini 3}}
    \put(73,  \pl){\textcolor{black}{ChatGPT 5.2}}
    \end{overpic}
    \vspace{0.5mm}
    \caption{\textbf{Compatibility with Alternative 2D Guidance Signals.} Our method is modular and while we use Flux~\cite{labs2025flux1kontextflowmatching} (left) for the results shown in the main paper, we also show our method with alternative 2D priors Gemini 3~\cite{team2023gemini} (middle) and ChatGPT 5.2~\cite{singh2025openai} (right). Our method is robust to different 2D guidance signals (shown in gray) producing plausible semantically aligned quads in each case.}
    \label{fig:diff-priors}
\end{figure}

\subsection{Properties and Robustness of \ourmethod{}.}
\label{subsec:robustness}
\noindent\textbf{Compatibility with Alternative 2D Guidance Signals.}\quad
For the main results in the paper, we use Flux~\cite{labs2025flux1kontextflowmatching} as our image generation model, however, our method is modular and robust to different 2D guidance signals. In \cref{fig:diff-priors}, we show results from our method using three different 2D text-to-image priors: Flux (left), Gemini 3~\cite{team2023gemini}) (middle), and ChatGPT 5.2~\cite{singh2025openai} (right). We observe that even with different priors, we still obtain meshes that align to salient features in the mesh.

\begin{figure}
    \centering
    \newcommand{\pl}{-4}
    \begin{overpic}[width=\linewidth]{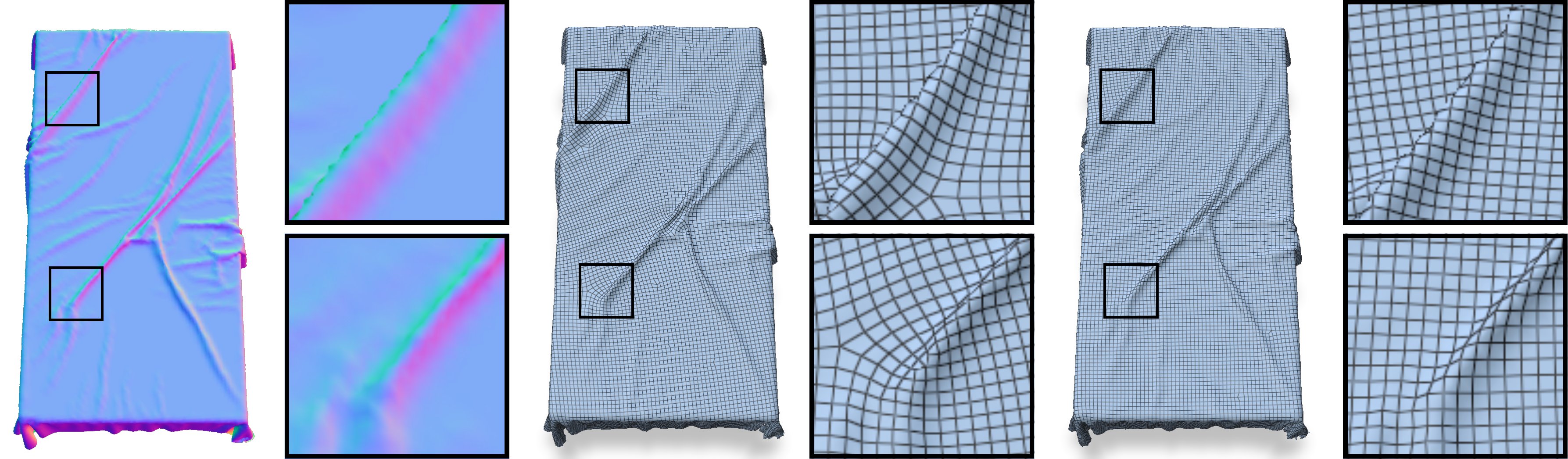}
        \put(10,  \pl){\textcolor{black}{Normals}}
        \put(37,  \pl){\textcolor{black}{Curvature-aligned}}
        \put(80,  \pl){\textcolor{black}{\textbf{Ours}}}
    \end{overpic}
    \vspace{-0.25cm}
    \caption{\textbf{Robustness to Surface Noise}. Our visually-guided method is robust to surface noise such as the wrinkles in the tablecloth. A baseline method using principal curvature constraints alongside a smoothness term aligns to these local features producing undesirable edge flow and introducing additional singularities.}
    \label{fig:wrinkles}
\end{figure}

\smallskip
\noindent\textbf{Robustness to Surface Noise.}\quad
Since our method is visually guided, it is robust to sharp variations in surface geometry. In \cref{fig:wrinkles}, we show a mesh of a picnic table covered by a wrinkled tablecloth. These wrinkles produce sharp ridges in the mesh geometry as indicated by the surface normals (left). Approaches that draw alignment from geometric features are very sensitive to such surface noise. We compare to a baseline implementation that replaces our visually-informed alignment constraints with constraints aligned to the principal curvature of the mesh, a commonly used geometric feature (middle). As expected, this setting aligns to the sharp ridges of the wrinkles which produces undesirable edge flow that ignores the global context and introduces additional singularities. In contrast, our method leverages the strong prior of 2D generative models which use the global context to understand that these wrinkles are only noise. Our guidance instead correctly treats the tablecloth as intrinsically flat and aligns the edge flow accordingly. This results in a more realistic meshing alignment and fewer singularities.

\begin{figure}
    \centering
    \includegraphics[width=\linewidth]{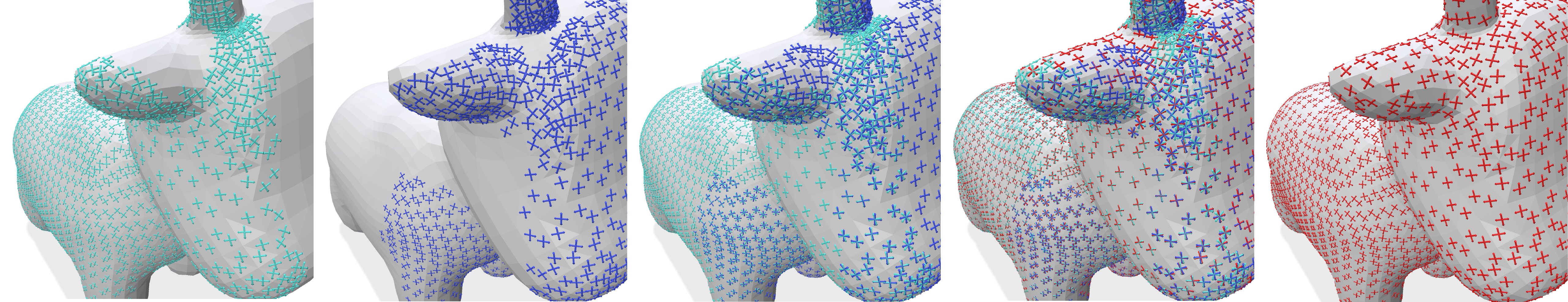}
    \caption{\textbf{Multi-view Interpolation}. Our approach smoothly interpolates per-view cross fields (cyan, blue) into a single smooth cross field for the entire mesh (red).}
    \label{fig:smoothness}
\end{figure}

\begin{figure}
    \centering
    \newcommand{\pl}{-3}
    \begin{overpic}[width=\linewidth]{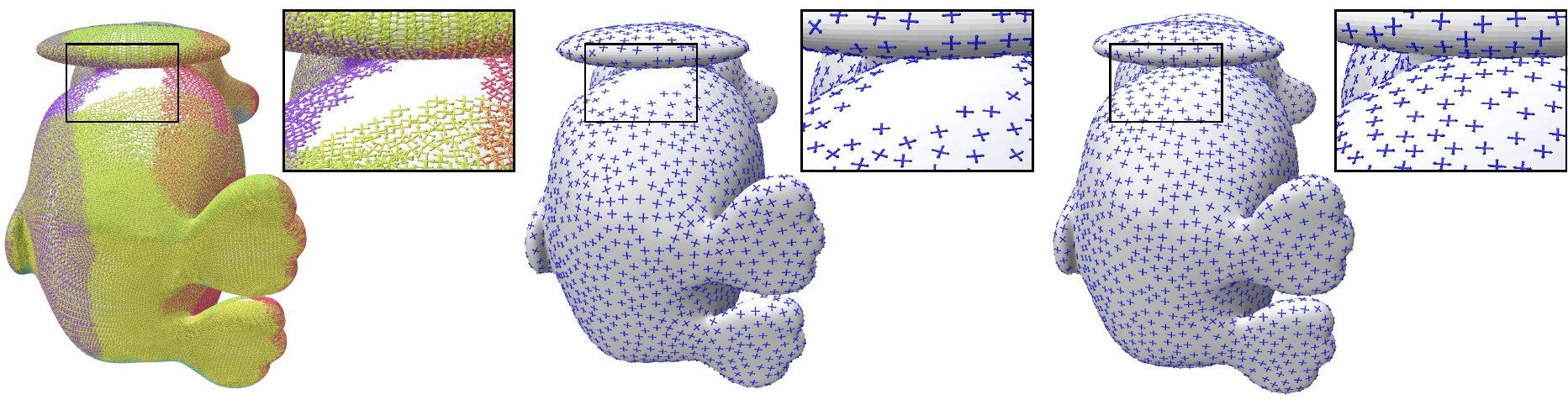}
        \put(1,  \pl){\textcolor{black}{Multi-view Crosses}}
        \put(37,  \pl){\textcolor{black}{Solve w/ $\lambda_s = 0$}}
        \put(72,  \pl){\textcolor{black}{Solve w/ $\lambda_s = 1$}}
    \end{overpic}
    \caption{\textbf{Interpolation to Non-visible Mesh Regions}. Our method robustly produces a smooth field even in regions that are not visible in the 2D views (\ie the area under the penguin's wing). Without the smoothness term ($\lambda_s = 0$), the resulting field has holes and our quad mesh extraction fails. With this term ($\lambda_s = 1$), we are able to smoothly interpolate the surrounding field to get meaningful values in the unseen region.}
    \label{fig:smoothness-term}
\end{figure}

\smallskip
\noindent\textbf{Smooth Interpolations to Non-Visible  Mesh Regions.}\quad
Our interpolations in \cref{subsec:per-view-crosses} and \cref{subsec:multi-view-crosses} smoothly interpolate directions over faces. We show in \cref{fig:smoothness} that we can take the per-view crosses from \textit{only} $2$ views (cyan, blue) such that many faces do not have an alignment signal and use our multi-view interpolation to produce a smooth cross field (red) that contains signal for all faces on the mesh. This property can be utilized to determine alignment in regions of the mesh that are not visible from the $6$ views used in our visual guidance and thus contain no alignment signal. In \cref{fig:smoothness-term} we demonstrate this in practice by running an ablation on the smoothness term in our multi-view solve. Given cross fields derived from all $6$ views of the penguin (left), removing the smoothness term from the multi-view solve ($\lambda_s=0$, middle) results in no alignment signal in the occluded region underneath the penguin's wing. Using our full method with the smoothness term ($\lambda_s=1$, right) interpolates the nearby, well-defined alignment values to produce smooth and plausible alignment even in this occluded region underneath the penguin's wing.

\begin{figure}
    \centering
    \newcommand{\pl}{-4}
    \begin{overpic}[width=\linewidth]{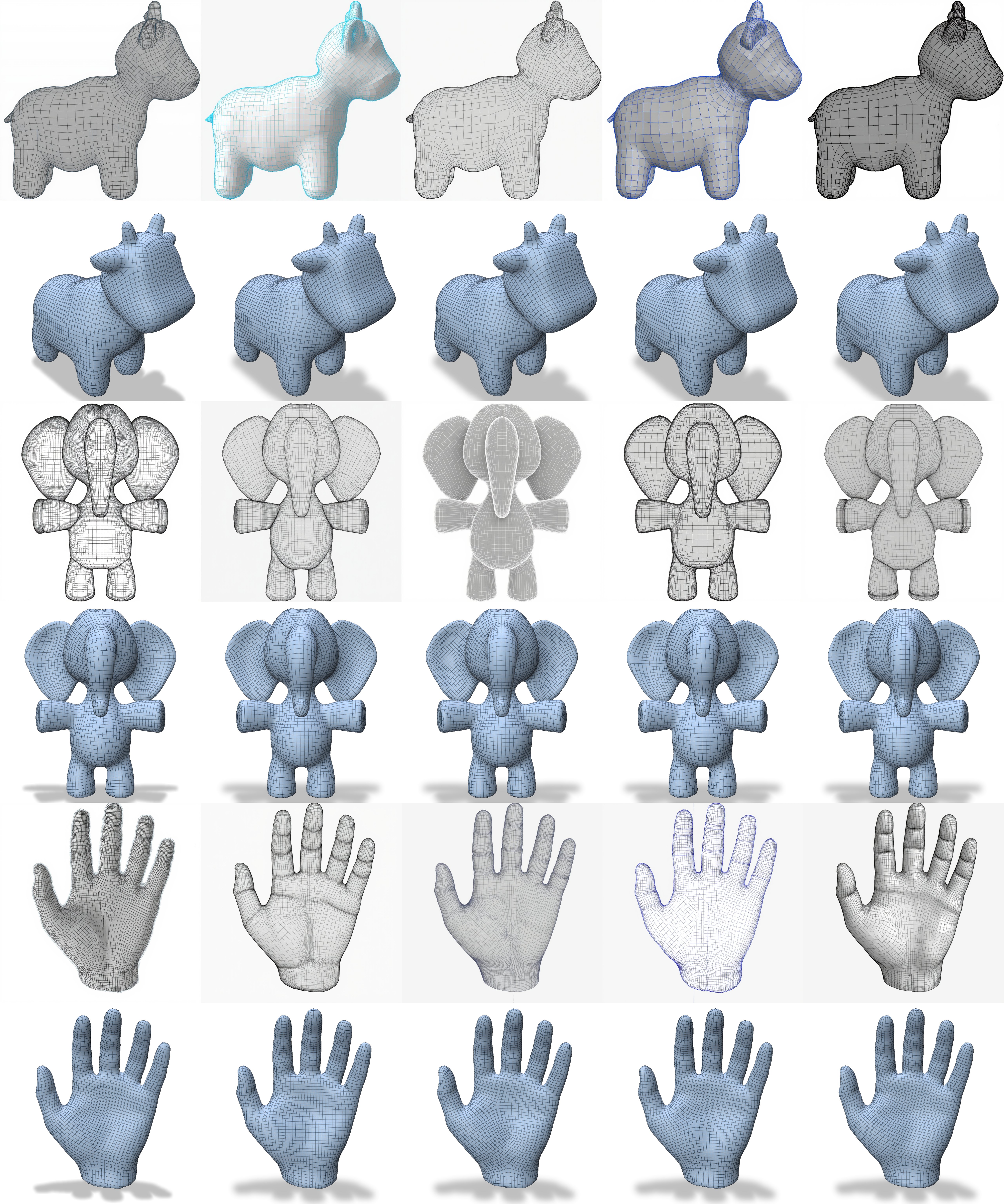}
        \put(4.5,  \pl){\textcolor{black}{Seed A}}
        \put(21.5,  \pl){\textcolor{black}{Seed B}}
        \put(38,  \pl){\textcolor{black}{Seed C}}
        \put(54.5,  \pl){\textcolor{black}{Seed D}}
        \put(71,  \pl){\textcolor{black}{Seed E}}
    \end{overpic}
    \vspace{1mm}
    \caption{\textbf{Robustness to Random Seeds and Model Variation}. While our 2D guidance model can output images with slight variations in color, scale, and noisy artifacts when using different seeds (odd rows), the high level alignment signal contained in these images is consistent and thus our method is able to produce similar quad meshes in each case (even rows).}
    \label{fig:seed}
\end{figure}
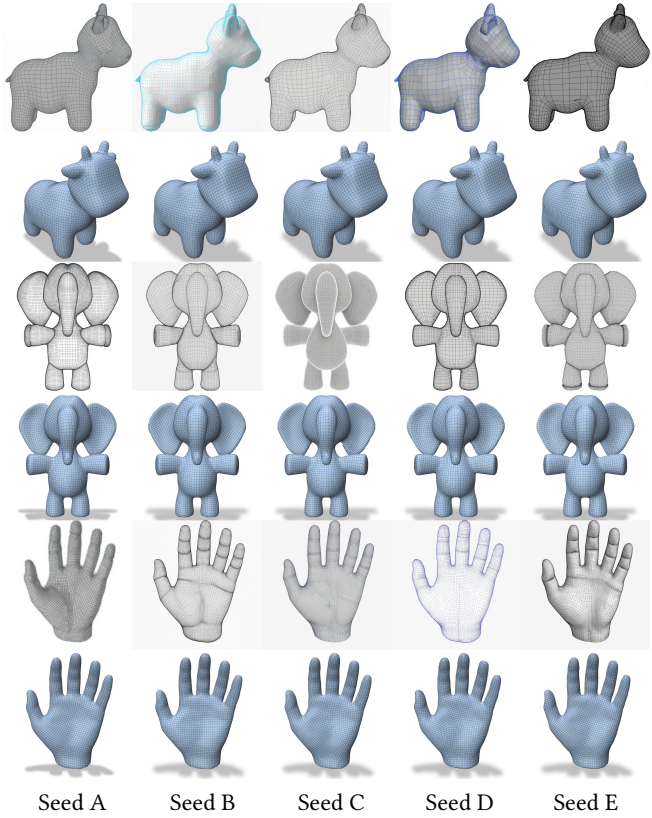

\smallskip
\noindent\textbf{Robustness to Random Seeds and Model Variation.}
Our method relies on pre-trained 2D generative models to guide our alignment. As such, the guidance images we generate inherit the non-determinism of these models. However, our two-stage interpolation is sufficiently robust to this noise and while our quad outputs do exhibit slight variations, the high level quad alignment is stable. In \cref{fig:seed}, we show the images generated using $5$ different seeds for $3$ different meshes. Across the different seeds, the generated guidance images exhibit variations in color, quad scale, and noisy artifacts. However, the majority of the edge flow alignment that we extract from these images is correct and using this noisy, but informative signal we are able to reconstruct a smooth, feature aligned field in each case. While specific details such as exact placement of singularities can vary slightly, our resulting quad meshes retain similar edge flow alignment across all $5$ seeds.

\begin{figure}
    \centering
    \includegraphics[width=\linewidth]{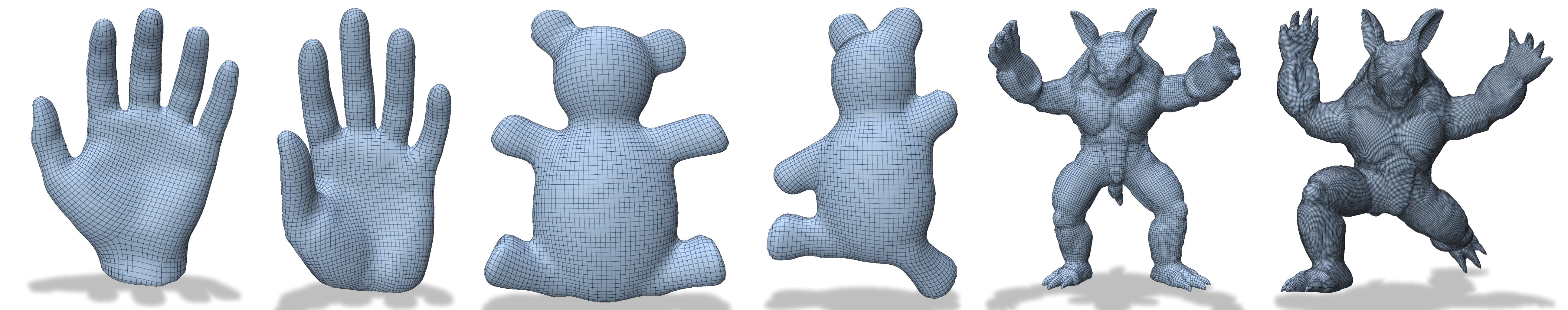}
    \caption{\textbf{Robustness to Pose and Deformations.} If we deform or alter the pose of our input mesh, we still produce quads that are aligned to the same high level features.\dale{Maybe need to rename to something along the lines of ``Robustness across variations of the same object''}}
    \label{fig:diff-pose}
\end{figure}

\smallskip
\noindent\textbf{Robustness to Pose and Deformations}\quad
In \cref{fig:diff-pose}, we apply our method to meshes representing the same object in different poses or with minor deformations and geometric differences. In these cases, \ourmethod{} produces quad meshes with similar high level edge flow, despite the differing mesh inputs. For example, the ears, arms, and legs of the teddy bear (middle) are all angled differently relative to the main torso in the two poses. Nonetheless, our method still produces alignment that follows the main axes of these appendages, regardless of orientation.

\begin{figure}
    \centering
    \includegraphics[width=\linewidth]{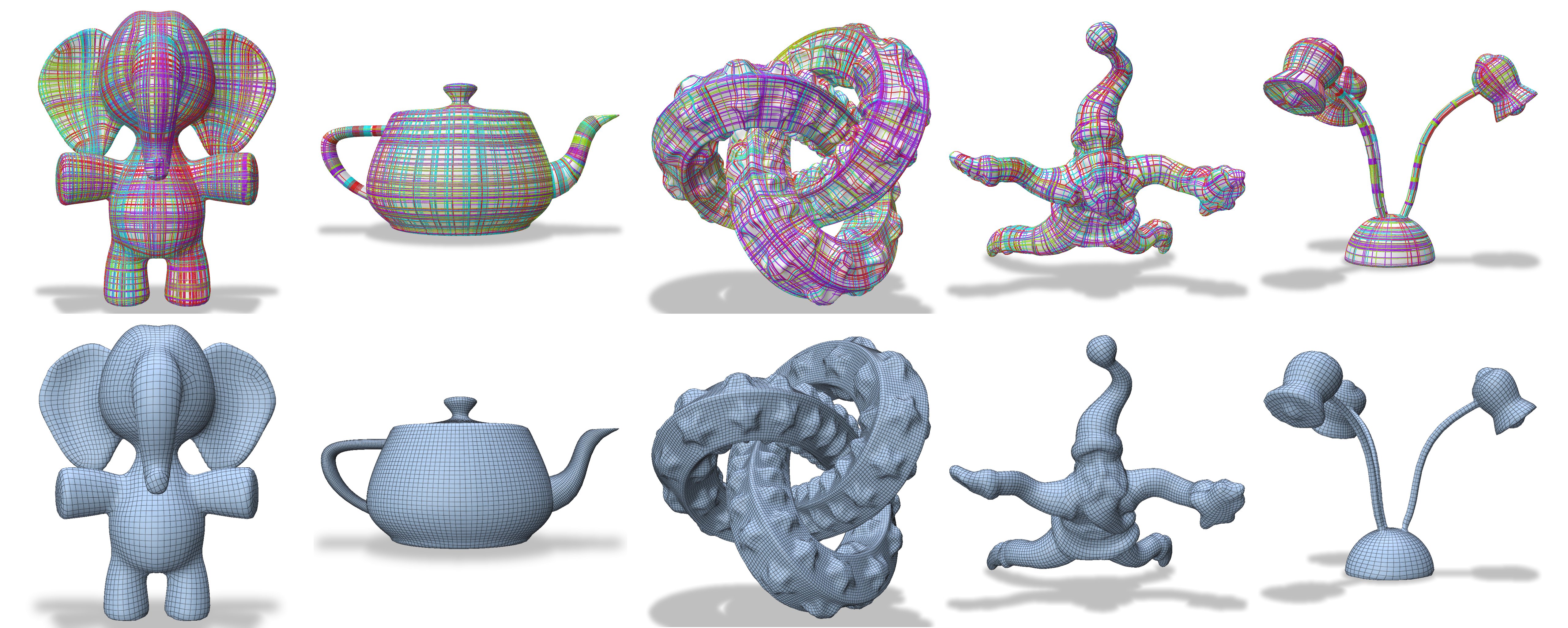}
    \caption{\textbf{Streamlines Visualization.} We visualize streamlines of our cross field traced across the mesh surface (top) and the corresponding quad meshes we extract (bottom).}
    \label{fig:streamlines}
\end{figure}

\begin{figure}
    \centering
    \newcommand{\pl}{-4}
    \newcommand{\snd}{-8}
    \includegraphics[width=\linewidth]{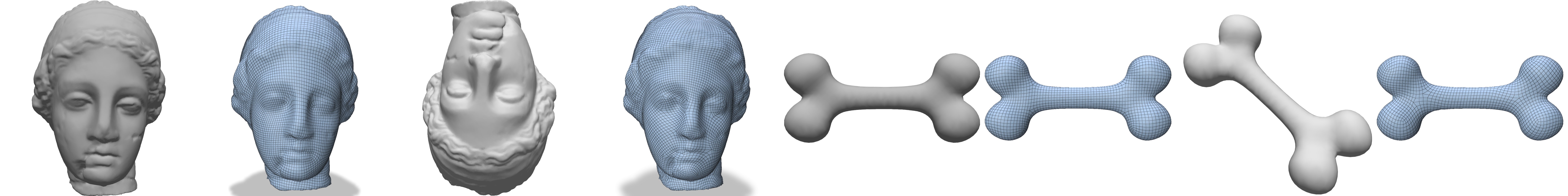}
    \caption{\textbf{Limitations}. Our method is not equivariant to rigid transformations. We run \ourmethod{} on the same mesh before and after a rigid transformation (gray) and show the resulting quad meshes (blue). While our results display similar alignment to high level features, the quad meshes themselves are not guaranteed to be identical.}
    \label{fig:limitations}
\end{figure}

\smallskip
\noindent\textbf{Limitations.}\quad
\ourmethod{} is not guaranteed to be equivariant to rigid transformations of the mesh. As shown in \cref{fig:limitations}, when we run our method on different rigid transformations, we obtain quad meshes that capture similar high level features, however, there are no guarantees that the quad meshes we produce in these cases are identical.

\section{Conclusion}
In this work, we presented \ourmethod{}, a technique for generating feature-aligned quad meshes from 2D visual input. Manually aligning quads requires substantial 3D experience and skill to use and automated methods for quad meshing rely mostly on geometry which is not sufficient for robust feature alignment. We overcome these challenges by leveraging the strong semantic and geometric understanding of text-to-image models. Key to our approach is our two-stage interpolation that smoothly converts the image space gradients to a per-face cross field. We demonstrate the effectiveness of our technique on a diverse set of input meshes showing improved feature alignment over existing methods. Furthermore, our visually-guided approach allows for unique applications such as texture-aligned quad meshing and alignment guidance from intuitive user-drawn 2D images. In the future, we aim to apply 2D priors to other 3D tasks beyond feature alignment such as semantically-aware deformation.

\section{Acknowledgments}

We thank the University of Chicago for providing the AI cluster resources, services, and the professional support of the technical staff. This work was also generously supported by gifts from Adobe Research, Snap Research, Google Research, BSF grant 2022363, the Bennett Family AI + Science Collaborative Research Program, and NSF grants 2304481, 2402894, 2140001, and 2335493. Finally, we would like to thank Nam Anh Dinh, Itai Lang, and the members of 3DL for their thorough and insightful feedback on our work.

\bibliographystyle{ACM-Reference-Format}
\bibliography{main}


\begin{thebibliography}{37}


\ifx \showCODEN    \undefined \def \showCODEN     #1{\unskip}     \fi
\ifx \showDOI      \undefined \def \showDOI       #1{#1}\fi
\ifx \showISBNx    \undefined \def \showISBNx     #1{\unskip}     \fi
\ifx \showISBNxiii \undefined \def \showISBNxiii  #1{\unskip}     \fi
\ifx \showISSN     \undefined \def \showISSN      #1{\unskip}     \fi
\ifx \showLCCN     \undefined \def \showLCCN      #1{\unskip}     \fi
\ifx \shownote     \undefined \def \shownote      #1{#1}          \fi
\ifx \showarticletitle \undefined \def \showarticletitle #1{#1}   \fi
\ifx \showURL      \undefined \def \showURL       {\relax}        \fi
\providecommand\bibfield[2]{#2}
\providecommand\bibinfo[2]{#2}
\providecommand\natexlab[1]{#1}
\providecommand\showeprint[2][]{arXiv:#2}

\bibitem[Azencot et~al\mbox{.}(2017)]%
        {azencot2017consistent}
\bibfield{author}{\bibinfo{person}{Omri Azencot}, \bibinfo{person}{Etienne Corman}, \bibinfo{person}{Mirela Ben-Chen}, {and} \bibinfo{person}{Maks Ovsjanikov}.} \bibinfo{year}{2017}\natexlab{}.
\newblock \showarticletitle{Consistent functional cross field design for mesh quadrangulation}.
\newblock \bibinfo{journal}{\emph{ACM Transactions on Graphics (TOG)}} \bibinfo{volume}{36}, \bibinfo{number}{4} (\bibinfo{year}{2017}), \bibinfo{pages}{1--13}.
\newblock


\bibitem[{Blender Studio}(2026)]%
        {BlenderStudio2026}
\bibfield{author}{\bibinfo{person}{{Blender Studio}}.} \bibinfo{year}{2026}\natexlab{}.
\newblock \bibinfo{booktitle}{\emph{Blender Fundamentals 4.5 LTS}}.
\newblock Blender Studio.
\newblock
\urldef\tempurl%
\url{https://studio.blender.org/training/blender-fundamentals-45-lts/}
\showURL{%
\tempurl}


\bibitem[Bommes et~al\mbox{.}(2012)]%
        {bommes2012state}
\bibfield{author}{\bibinfo{person}{David Bommes}, \bibinfo{person}{Bruno L{\'e}vy}, \bibinfo{person}{Nico Pietroni}, \bibinfo{person}{Enrico Puppo}, \bibinfo{person}{Claudio Silva}, \bibinfo{person}{Marco Tarini}, {and} \bibinfo{person}{Denis Zorin}.} \bibinfo{year}{2012}\natexlab{}.
\newblock \showarticletitle{State of the art in quad meshing}. In \bibinfo{booktitle}{\emph{Eurographics-33rd Annual Conference of the European Association for Computer Graphics-2012}}.
\newblock


\bibitem[Bommes et~al\mbox{.}(2013)]%
        {bommes2013quad}
\bibfield{author}{\bibinfo{person}{David Bommes}, \bibinfo{person}{Bruno L{\'e}vy}, \bibinfo{person}{Nico Pietroni}, \bibinfo{person}{Enrico Puppo}, \bibinfo{person}{Claudio Silva}, \bibinfo{person}{Marco Tarini}, {and} \bibinfo{person}{Denis Zorin}.} \bibinfo{year}{2013}\natexlab{}.
\newblock \showarticletitle{Quad-mesh generation and processing: A survey}. In \bibinfo{booktitle}{\emph{Computer graphics forum}}, Vol.~\bibinfo{volume}{32}. Wiley Online Library, \bibinfo{pages}{51--76}.
\newblock


\bibitem[Bommes et~al\mbox{.}(2009a)]%
        {10.1145/1531326.1531383}
\bibfield{author}{\bibinfo{person}{David Bommes}, \bibinfo{person}{Henrik Zimmer}, {and} \bibinfo{person}{Leif Kobbelt}.} \bibinfo{year}{2009}\natexlab{a}.
\newblock \showarticletitle{Mixed-integer quadrangulation}.
\newblock \bibinfo{journal}{\emph{ACM Trans. Graph.}} \bibinfo{volume}{28}, \bibinfo{number}{3}, Article \bibinfo{articleno}{77} (\bibinfo{date}{July} \bibinfo{year}{2009}), \bibinfo{numpages}{10}~pages.
\newblock
\showISSN{0730-0301}
\urldef\tempurl%
\url{https://doi.org/10.1145/1531326.1531383}
\showDOI{\tempurl}


\bibitem[Bommes et~al\mbox{.}(2009b)]%
        {bommes2009mixed}
\bibfield{author}{\bibinfo{person}{David Bommes}, \bibinfo{person}{Henrik Zimmer}, {and} \bibinfo{person}{Leif Kobbelt}.} \bibinfo{year}{2009}\natexlab{b}.
\newblock \showarticletitle{Mixed-integer quadrangulation}.
\newblock \bibinfo{journal}{\emph{ACM transactions on graphics (TOG)}} \bibinfo{volume}{28}, \bibinfo{number}{3} (\bibinfo{year}{2009}), \bibinfo{pages}{1--10}.
\newblock


\bibitem[Brandt et~al\mbox{.}(2018)]%
        {brandt2018modeling}
\bibfield{author}{\bibinfo{person}{Christopher Brandt}, \bibinfo{person}{Leonardo Scandolo}, \bibinfo{person}{Elmar Eisemann}, {and} \bibinfo{person}{Klaus Hildebrandt}.} \bibinfo{year}{2018}\natexlab{}.
\newblock \showarticletitle{Modeling n-symmetry vector fields using higher-order energies}.
\newblock \bibinfo{journal}{\emph{ACM Transactions on Graphics (TOG)}} \bibinfo{volume}{37}, \bibinfo{number}{2} (\bibinfo{year}{2018}), \bibinfo{pages}{1--18}.
\newblock


\bibitem[Crane et~al\mbox{.}(2010)]%
        {Crane:2010:TCD}
\bibfield{author}{\bibinfo{person}{Keenan Crane}, \bibinfo{person}{Mathieu Desbrun}, {and} \bibinfo{person}{Peter Schr\"{o}der}.} \bibinfo{year}{2010}\natexlab{}.
\newblock \showarticletitle{Trivial Connections on Discrete Surfaces}.
\newblock \bibinfo{journal}{\emph{Computer Graphics Forum (SGP)}} \bibinfo{volume}{29}, \bibinfo{number}{5} (\bibinfo{year}{2010}), \bibinfo{pages}{1525--1533}.
\newblock


\bibitem[Diamanti et~al\mbox{.}(2015)]%
        {10.1145/2766906}
\bibfield{author}{\bibinfo{person}{Olga Diamanti}, \bibinfo{person}{Amir Vaxman}, \bibinfo{person}{Daniele Panozzo}, {and} \bibinfo{person}{Olga Sorkine-Hornung}.} \bibinfo{year}{2015}\natexlab{}.
\newblock \showarticletitle{Integrable PolyVector fields}.
\newblock \bibinfo{journal}{\emph{ACM Trans. Graph.}} \bibinfo{volume}{34}, \bibinfo{number}{4}, Article \bibinfo{articleno}{38} (\bibinfo{date}{July} \bibinfo{year}{2015}), \bibinfo{numpages}{12}~pages.
\newblock
\showISSN{0730-0301}
\urldef\tempurl%
\url{https://doi.org/10.1145/2766906}
\showDOI{\tempurl}


\bibitem[Dielen et~al\mbox{.}(2021)]%
        {https://doi.org/10.1111/cgf.14366}
\bibfield{author}{\bibinfo{person}{Alexander Dielen}, \bibinfo{person}{Isaak Lim}, \bibinfo{person}{Max Lyon}, {and} \bibinfo{person}{Leif Kobbelt}.} \bibinfo{year}{2021}\natexlab{}.
\newblock \showarticletitle{Learning Direction Fields for Quad Mesh Generation}.
\newblock \bibinfo{journal}{\emph{Computer Graphics Forum}} \bibinfo{volume}{40}, \bibinfo{number}{5} (\bibinfo{year}{2021}), \bibinfo{pages}{181--191}.
\newblock
\urldef\tempurl%
\url{https://doi.org/10.1111/cgf.14366}
\showDOI{\tempurl}
\showeprint{https://onlinelibrary.wiley.com/doi/pdf/10.1111/cgf.14366}


\bibitem[Dong et~al\mbox{.}(2025a)]%
        {dong2025crossgen}
\bibfield{author}{\bibinfo{person}{Qiujie Dong}, \bibinfo{person}{Jiepeng Wang}, \bibinfo{person}{Rui Xu}, \bibinfo{person}{Cheng Lin}, \bibinfo{person}{Yuan Liu}, \bibinfo{person}{Shiqing Xin}, \bibinfo{person}{Zichun Zhong}, \bibinfo{person}{Xin Li}, \bibinfo{person}{Changhe Tu}, \bibinfo{person}{Taku Komura}, {et~al\mbox{.}}} \bibinfo{year}{2025}\natexlab{a}.
\newblock \showarticletitle{CrossGen: Learning and Generating Cross Fields for Quad Meshing}.
\newblock \bibinfo{journal}{\emph{ACM Transactions on Graphics (TOG)}} \bibinfo{volume}{44}, \bibinfo{number}{6} (\bibinfo{year}{2025}), \bibinfo{pages}{1--15}.
\newblock


\bibitem[Dong et~al\mbox{.}(2025b)]%
        {dong2025neurcrossneuralapproachcomputing}
\bibfield{author}{\bibinfo{person}{Qiujie Dong}, \bibinfo{person}{Huibiao Wen}, \bibinfo{person}{Rui Xu}, \bibinfo{person}{Shuangmin Chen}, \bibinfo{person}{Jiaran Zhou}, \bibinfo{person}{Shiqing Xin}, \bibinfo{person}{Changhe Tu}, \bibinfo{person}{Taku Komura}, {and} \bibinfo{person}{Wenping Wang}.} \bibinfo{year}{2025}\natexlab{b}.
\newblock \bibinfo{title}{NeurCross: A Neural Approach to Computing Cross Fields for Quad Mesh Generation}.
\newblock
\newblock
\showeprint[arxiv]{2405.13745}~[cs.CV]
\urldef\tempurl%
\url{https://arxiv.org/abs/2405.13745}
\showURL{%
\tempurl}


\bibitem[Du et~al\mbox{.}(2023)]%
        {du2023generative}
\bibfield{author}{\bibinfo{person}{Xiaodan Du}, \bibinfo{person}{Nicholas Kolkin}, \bibinfo{person}{Greg Shakhnarovich}, {and} \bibinfo{person}{Anand Bhattad}.} \bibinfo{year}{2023}\natexlab{}.
\newblock \showarticletitle{Generative models: What do they know? do they know things? let's find out!}
\newblock \bibinfo{journal}{\emph{arXiv preprint arXiv:2311.17137}} (\bibinfo{year}{2023}).
\newblock


\bibitem[Ebke et~al\mbox{.}(2013)]%
        {Ebke:2013:QRQ:2508363.2508372}
\bibfield{author}{\bibinfo{person}{Hans-Christian Ebke}, \bibinfo{person}{David Bommes}, \bibinfo{person}{Marcel Campen}, {and} \bibinfo{person}{Leif Kobbelt}.} \bibinfo{year}{2013}\natexlab{}.
\newblock \showarticletitle{{QE}x: Robust Quad Mesh Extraction}.
\newblock \bibinfo{journal}{\emph{ACM Trans. Graph.}} \bibinfo{volume}{32}, \bibinfo{number}{6}, Article \bibinfo{articleno}{168} (\bibinfo{date}{Nov.} \bibinfo{year}{2013}), \bibinfo{numpages}{10}~pages.
\newblock
\showISSN{0730-0301}
\urldef\tempurl%
\url{https://doi.org/10.1145/2508363.2508372}
\showDOI{\tempurl}


\bibitem[Girshick et~al\mbox{.}(2000)]%
        {girshick2000line}
\bibfield{author}{\bibinfo{person}{Ahna Girshick}, \bibinfo{person}{Victoria Interrante}, \bibinfo{person}{Steven Haker}, {and} \bibinfo{person}{Todd Lemoine}.} \bibinfo{year}{2000}\natexlab{}.
\newblock \showarticletitle{Line direction matters: An argument for the use of principal directions in 3D line drawings}. In \bibinfo{booktitle}{\emph{Proceedings of the 1st International Symposium on Non-photorealistic Animation and Rendering}}. \bibinfo{pages}{43--52}.
\newblock


\bibitem[Hao et~al\mbox{.}(2024)]%
        {hao2024meshtronhighfidelityartistlike3d}
\bibfield{author}{\bibinfo{person}{Zekun Hao}, \bibinfo{person}{David~W. Romero}, \bibinfo{person}{Tsung-Yi Lin}, {and} \bibinfo{person}{Ming-Yu Liu}.} \bibinfo{year}{2024}\natexlab{}.
\newblock \bibinfo{title}{Meshtron: High-Fidelity, Artist-Like 3D Mesh Generation at Scale}.
\newblock
\newblock
\showeprint[arxiv]{2412.09548}~[cs.GR]
\urldef\tempurl%
\url{https://arxiv.org/abs/2412.09548}
\showURL{%
\tempurl}


\bibitem[Huang et~al\mbox{.}(2018)]%
        {10.1111:cgf.13498}
\bibfield{author}{\bibinfo{person}{Jingwei Huang}, \bibinfo{person}{Yichao Zhou}, \bibinfo{person}{Matthias Niessner}, \bibinfo{person}{Jonathan~Richard Shewchuk}, {and} \bibinfo{person}{Leonidas~J. Guibas}.} \bibinfo{year}{2018}\natexlab{}.
\newblock \showarticletitle{{QuadriFlow: A Scalable and Robust Method for Quadrangulation}}.
\newblock \bibinfo{journal}{\emph{Computer Graphics Forum}} (\bibinfo{year}{2018}).
\newblock
\showISSN{1467-8659}
\urldef\tempurl%
\url{https://doi.org/10.1111/cgf.13498}
\showDOI{\tempurl}


\bibitem[Jacobson et~al\mbox{.}(2018)]%
        {libigl}
\bibfield{author}{\bibinfo{person}{Alec Jacobson}, \bibinfo{person}{Daniele Panozzo}, {et~al\mbox{.}}} \bibinfo{year}{2018}\natexlab{}.
\newblock \bibinfo{title}{{libigl}: A simple {C++} geometry processing library}.
\newblock
\newblock
\newblock
\shownote{https://libigl.github.io/}.


\bibitem[Jakob et~al\mbox{.}(2015)]%
        {Jakob2015Instant}
\bibfield{author}{\bibinfo{person}{Wenzel Jakob}, \bibinfo{person}{Marco Tarini}, \bibinfo{person}{Daniele Panozzo}, {and} \bibinfo{person}{Olga Sorkine-Hornung}.} \bibinfo{year}{2015}\natexlab{}.
\newblock \showarticletitle{Instant Field-Aligned Meshes}.
\newblock \bibinfo{journal}{\emph{ACM Transactions on Graphics (Proceedings of SIGGRAPH ASIA)}} \bibinfo{volume}{34}, \bibinfo{number}{6} (\bibinfo{date}{Nov.} \bibinfo{year}{2015}).
\newblock
\urldef\tempurl%
\url{https://doi.org/10.1145/2816795.2818078}
\showDOI{\tempurl}


\bibitem[Kn\"{o}ppel et~al\mbox{.}(2013)]%
        {10.1145/2461912.2462005}
\bibfield{author}{\bibinfo{person}{Felix Kn\"{o}ppel}, \bibinfo{person}{Keenan Crane}, \bibinfo{person}{Ulrich Pinkall}, {and} \bibinfo{person}{Peter Schr\"{o}der}.} \bibinfo{year}{2013}\natexlab{}.
\newblock \showarticletitle{Globally optimal direction fields}.
\newblock \bibinfo{journal}{\emph{ACM Trans. Graph.}} \bibinfo{volume}{32}, \bibinfo{number}{4}, Article \bibinfo{articleno}{59} (\bibinfo{date}{July} \bibinfo{year}{2013}), \bibinfo{numpages}{10}~pages.
\newblock
\showISSN{0730-0301}
\urldef\tempurl%
\url{https://doi.org/10.1145/2461912.2462005}
\showDOI{\tempurl}


\bibitem[Labs et~al\mbox{.}(2025)]%
        {labs2025flux1kontextflowmatching}
\bibfield{author}{\bibinfo{person}{Black~Forest Labs}, \bibinfo{person}{Stephen Batifol}, \bibinfo{person}{Andreas Blattmann}, \bibinfo{person}{Frederic Boesel}, \bibinfo{person}{Saksham Consul}, \bibinfo{person}{Cyril Diagne}, \bibinfo{person}{Tim Dockhorn}, \bibinfo{person}{Jack English}, \bibinfo{person}{Zion English}, \bibinfo{person}{Patrick Esser}, \bibinfo{person}{Sumith Kulal}, \bibinfo{person}{Kyle Lacey}, \bibinfo{person}{Yam Levi}, \bibinfo{person}{Cheng Li}, \bibinfo{person}{Dominik Lorenz}, \bibinfo{person}{Jonas Müller}, \bibinfo{person}{Dustin Podell}, \bibinfo{person}{Robin Rombach}, \bibinfo{person}{Harry Saini}, \bibinfo{person}{Axel Sauer}, {and} \bibinfo{person}{Luke Smith}.} \bibinfo{year}{2025}\natexlab{}.
\newblock \bibinfo{title}{FLUX.1 Kontext: Flow Matching for In-Context Image Generation and Editing in Latent Space}.
\newblock
\newblock
\showeprint[arxiv]{2506.15742}~[cs.GR]
\urldef\tempurl%
\url{https://arxiv.org/abs/2506.15742}
\showURL{%
\tempurl}


\bibitem[Li et~al\mbox{.}(2025)]%
        {Li_2025}
\bibfield{author}{\bibinfo{person}{Zezeng Li}, \bibinfo{person}{Zhihui Qi}, \bibinfo{person}{Weimin Wang}, \bibinfo{person}{Ziliang Wang}, \bibinfo{person}{Junyi Duan}, {and} \bibinfo{person}{Na Lei}.} \bibinfo{year}{2025}\natexlab{}.
\newblock \showarticletitle{Point2Quad: Generating Quad Meshes From Point Clouds via Face Prediction}.
\newblock \bibinfo{journal}{\emph{IEEE Transactions on Circuits and Systems for Video Technology}} \bibinfo{volume}{35}, \bibinfo{number}{9} (\bibinfo{date}{Sept.} \bibinfo{year}{2025}), \bibinfo{pages}{8586–8597}.
\newblock
\showISSN{1558-2205}
\urldef\tempurl%
\url{https://doi.org/10.1109/tcsvt.2025.3556130}
\showDOI{\tempurl}


\bibitem[Liu et~al\mbox{.}(2025)]%
        {NeuFrameQ-25}
\bibfield{author}{\bibinfo{person}{Ying-Tian Liu}, \bibinfo{person}{Jiajun Li}, \bibinfo{person}{Yu-Tao Liu}, \bibinfo{person}{Xin Yu}, \bibinfo{person}{Yuan-Chen Guo}, \bibinfo{person}{Yan-Pei Cao}, \bibinfo{person}{Ding Liang}, \bibinfo{person}{Ariel Shamir}, {and} \bibinfo{person}{Song-Hai Zhang}.} \bibinfo{year}{2025}\natexlab{}.
\newblock \showarticletitle{NeuFrameQ: Neural Frame Fields for Scalable and Generalizable Anisotropic Quadrangulation}. In \bibinfo{booktitle}{\emph{Proceedings of the International Conference on Computer Vision, ICCV}}. \bibinfo{pages}{accepted}.
\newblock


\bibitem[Meekes and Vaxman(2021)]%
        {meekes2021unconventional}
\bibfield{author}{\bibinfo{person}{Merel Meekes} {and} \bibinfo{person}{Amir Vaxman}.} \bibinfo{year}{2021}\natexlab{}.
\newblock \showarticletitle{Unconventional patterns on surfaces}.
\newblock \bibinfo{journal}{\emph{ACM Transactions on Graphics (TOG)}} \bibinfo{volume}{40}, \bibinfo{number}{4} (\bibinfo{year}{2021}), \bibinfo{pages}{1--16}.
\newblock


\bibitem[Panozzo et~al\mbox{.}(2014)]%
        {Panozzo:2014}
\bibfield{author}{\bibinfo{person}{Daniele Panozzo}, \bibinfo{person}{Enrico Puppo}, \bibinfo{person}{Marco Tarini}, {and} \bibinfo{person}{Olga Sorkine-Hornung}.} \bibinfo{year}{2014}\natexlab{}.
\newblock \showarticletitle{Frame Fields: Anisotropic and Non-Orthogonal Cross Fields}.
\newblock \bibinfo{journal}{\emph{ACM Transactions on Graphics (proceedings of ACM SIGGRAPH)}} \bibinfo{volume}{33}, \bibinfo{number}{4} (\bibinfo{year}{2014}), \bibinfo{pages}{134:1--134:11}.
\newblock


\bibitem[Pietroni et~al\mbox{.}(2021)]%
        {10.1145/3450626.3459941}
\bibfield{author}{\bibinfo{person}{Nico Pietroni}, \bibinfo{person}{Stefano Nuvoli}, \bibinfo{person}{Thomas Alderighi}, \bibinfo{person}{Paolo Cignoni}, {and} \bibinfo{person}{Marco Tarini}.} \bibinfo{year}{2021}\natexlab{}.
\newblock \showarticletitle{Reliable feature-line driven quad-remeshing}.
\newblock \bibinfo{journal}{\emph{ACM Trans. Graph.}} \bibinfo{volume}{40}, \bibinfo{number}{4}, Article \bibinfo{articleno}{155} (\bibinfo{date}{July} \bibinfo{year}{2021}), \bibinfo{numpages}{17}~pages.
\newblock
\showISSN{0730-0301}
\urldef\tempurl%
\url{https://doi.org/10.1145/3450626.3459941}
\showDOI{\tempurl}


\bibitem[Ray et~al\mbox{.}(2008)]%
        {10.1145/1356682.1356683}
\bibfield{author}{\bibinfo{person}{Nicolas Ray}, \bibinfo{person}{Bruno Vallet}, \bibinfo{person}{Wan~Chiu Li}, {and} \bibinfo{person}{Bruno L\'{e}vy}.} \bibinfo{year}{2008}\natexlab{}.
\newblock \showarticletitle{N-symmetry direction field design}.
\newblock \bibinfo{journal}{\emph{ACM Trans. Graph.}} \bibinfo{volume}{27}, \bibinfo{number}{2}, Article \bibinfo{articleno}{10} (\bibinfo{date}{May} \bibinfo{year}{2008}), \bibinfo{numpages}{13}~pages.
\newblock
\showISSN{0730-0301}
\urldef\tempurl%
\url{https://doi.org/10.1145/1356682.1356683}
\showDOI{\tempurl}


\bibitem[Scharr(2000)]%
        {Scharr2000OptimaleOI}
\bibfield{author}{\bibinfo{person}{Hanno Scharr}.} \bibinfo{year}{2000}\natexlab{}.
\newblock \showarticletitle{Optimale Operatoren in der Digitalen Bildverarbeitung}.
\newblock
\urldef\tempurl%
\url{https://api.semanticscholar.org/CorpusID:170511209}
\showURL{%
\tempurl}


\bibitem[Singh et~al\mbox{.}(2025)]%
        {singh2025openai}
\bibfield{author}{\bibinfo{person}{Aaditya Singh}, \bibinfo{person}{Adam Fry}, \bibinfo{person}{Adam Perelman}, \bibinfo{person}{Adam Tart}, \bibinfo{person}{Adi Ganesh}, \bibinfo{person}{Ahmed El-Kishky}, \bibinfo{person}{Aidan McLaughlin}, \bibinfo{person}{Aiden Low}, \bibinfo{person}{AJ Ostrow}, \bibinfo{person}{Akhila Ananthram}, {et~al\mbox{.}}} \bibinfo{year}{2025}\natexlab{}.
\newblock \showarticletitle{Openai gpt-5 system card}.
\newblock \bibinfo{journal}{\emph{arXiv preprint arXiv:2601.03267}} (\bibinfo{year}{2025}).
\newblock


\bibitem[Team et~al\mbox{.}(2023)]%
        {team2023gemini}
\bibfield{author}{\bibinfo{person}{Gemini Team}, \bibinfo{person}{Rohan Anil}, \bibinfo{person}{Sebastian Borgeaud}, \bibinfo{person}{Jean-Baptiste Alayrac}, \bibinfo{person}{Jiahui Yu}, \bibinfo{person}{Radu Soricut}, \bibinfo{person}{Johan Schalkwyk}, \bibinfo{person}{Andrew~M Dai}, \bibinfo{person}{Anja Hauth}, \bibinfo{person}{Katie Millican}, {et~al\mbox{.}}} \bibinfo{year}{2023}\natexlab{}.
\newblock \showarticletitle{Gemini: a family of highly capable multimodal models}.
\newblock \bibinfo{journal}{\emph{arXiv preprint arXiv:2312.11805}} (\bibinfo{year}{2023}).
\newblock


\bibitem[TurboSquid(2021)]%
        {turbosquid}
\bibfield{author}{\bibinfo{person}{TurboSquid}.} \bibinfo{year}{2021}\natexlab{}.
\newblock \bibinfo{title}{TurboSquid 3D Model Repository}.
\newblock
\newblock
\newblock
\shownote{https://www.turbosquid.com/}.


\bibitem[van Kaick et~al\mbox{.}(2011)]%
        {coseg_2011}
\bibfield{author}{\bibinfo{person}{Oliver van Kaick}, \bibinfo{person}{Andrea Tagliasacchi}, \bibinfo{person}{Oana Sidi}, \bibinfo{person}{Hao Zhang}, \bibinfo{person}{Daniel Cohen-Or}, \bibinfo{person}{Lior Wolf}, {and} \bibinfo{person}{Ghassan Hamarneh}.} \bibinfo{year}{2011}\natexlab{}.
\newblock \showarticletitle{Prior knowledge for part correspondence}.
\newblock \bibinfo{journal}{\emph{Computer Graphics Forum}} \bibinfo{volume}{30}, \bibinfo{number}{2} (\bibinfo{year}{2011}), \bibinfo{pages}{553–562}.
\newblock
\urldef\tempurl%
\url{https://doi.org/10.1111/j.1467-8659.2011.01893.x}
\showDOI{\tempurl}


\bibitem[Vaxman et~al\mbox{.}({[n.\,d.]})]%
        {Directional}
\bibfield{author}{\bibinfo{person}{Amir Vaxman} {et~al\mbox{.}}} \bibinfo{year}{[n.\,d.]}\natexlab{}.
\newblock \bibinfo{title}{Directional: A library for Directional Field Synthesis, Design, and Processing}.
\newblock
\newblock
\urldef\tempurl%
\url{https://doi.org/10.5281/zenodo.3338174}
\showDOI{\tempurl}


\bibitem[Vaxman et~al\mbox{.}(2017)]%
        {10.1145/3084873.3084921}
\bibfield{author}{\bibinfo{person}{Amir Vaxman}, \bibinfo{person}{Marcel Campen}, \bibinfo{person}{Olga Diamanti}, \bibinfo{person}{David Bommes}, \bibinfo{person}{Klaus Hildebrandt}, \bibinfo{person}{Mirela Ben-Chen Technion}, {and} \bibinfo{person}{Daniele Panozzo}.} \bibinfo{year}{2017}\natexlab{}.
\newblock \showarticletitle{Directional field synthesis, design, and processing}. In \bibinfo{booktitle}{\emph{ACM SIGGRAPH 2017 Courses}} (Los Angeles, California) \emph{(\bibinfo{series}{SIGGRAPH '17})}. \bibinfo{publisher}{Association for Computing Machinery}, \bibinfo{address}{New York, NY, USA}, Article \bibinfo{articleno}{12}, \bibinfo{numpages}{30}~pages.
\newblock
\showISBNx{9781450350143}
\urldef\tempurl%
\url{https://doi.org/10.1145/3084873.3084921}
\showDOI{\tempurl}


\bibitem[Viertel and Osting(2018)]%
        {viertel2018approachquadmeshingbased}
\bibfield{author}{\bibinfo{person}{Ryan Viertel} {and} \bibinfo{person}{Braxton Osting}.} \bibinfo{year}{2018}\natexlab{}.
\newblock \bibinfo{title}{An Approach to Quad Meshing Based on Harmonic Cross-Valued Maps and the Ginzburg-Landau Theory}.
\newblock
\newblock
\showeprint[arxiv]{1708.02316}~[cs.CG]
\urldef\tempurl%
\url{https://arxiv.org/abs/1708.02316}
\showURL{%
\tempurl}


\bibitem[Zhang et~al\mbox{.}(2023)]%
        {zhang2023adding}
\bibfield{author}{\bibinfo{person}{Lvmin Zhang}, \bibinfo{person}{Anyi Rao}, {and} \bibinfo{person}{Maneesh Agrawala}.} \bibinfo{year}{2023}\natexlab{}.
\newblock \showarticletitle{Adding conditional control to text-to-image diffusion models}. In \bibinfo{booktitle}{\emph{Proceedings of the IEEE/CVF international conference on computer vision}}. \bibinfo{pages}{3836--3847}.
\newblock


\bibitem[Zhang et~al\mbox{.}(2020)]%
        {zhang2020octahedralframesfeaturealignedcrossfields}
\bibfield{author}{\bibinfo{person}{Paul Zhang}, \bibinfo{person}{Josh Vekhter}, \bibinfo{person}{Edward Chien}, \bibinfo{person}{David Bommes}, \bibinfo{person}{Etienne Vouga}, {and} \bibinfo{person}{Justin Solomon}.} \bibinfo{year}{2020}\natexlab{}.
\newblock \bibinfo{title}{Octahedral Frames for Feature-Aligned Cross-Fields}.
\newblock
\newblock
\showeprint[arxiv]{2007.09740}~[cs.GR]
\urldef\tempurl%
\url{https://arxiv.org/abs/2007.09740}
\showURL{%
\tempurl}


\end{thebibliography}

\clearpage
\maketitleappendix
\appendix

\section{Optional Settings}
We outlines several additional settings and techniques that we do not use in our main method, but could be useful situationally and provide further control to the user.

\begin{figure}[b]
    \centering
    \newcommand{\pl}{-4}
    \begin{overpic}[width=\linewidth]{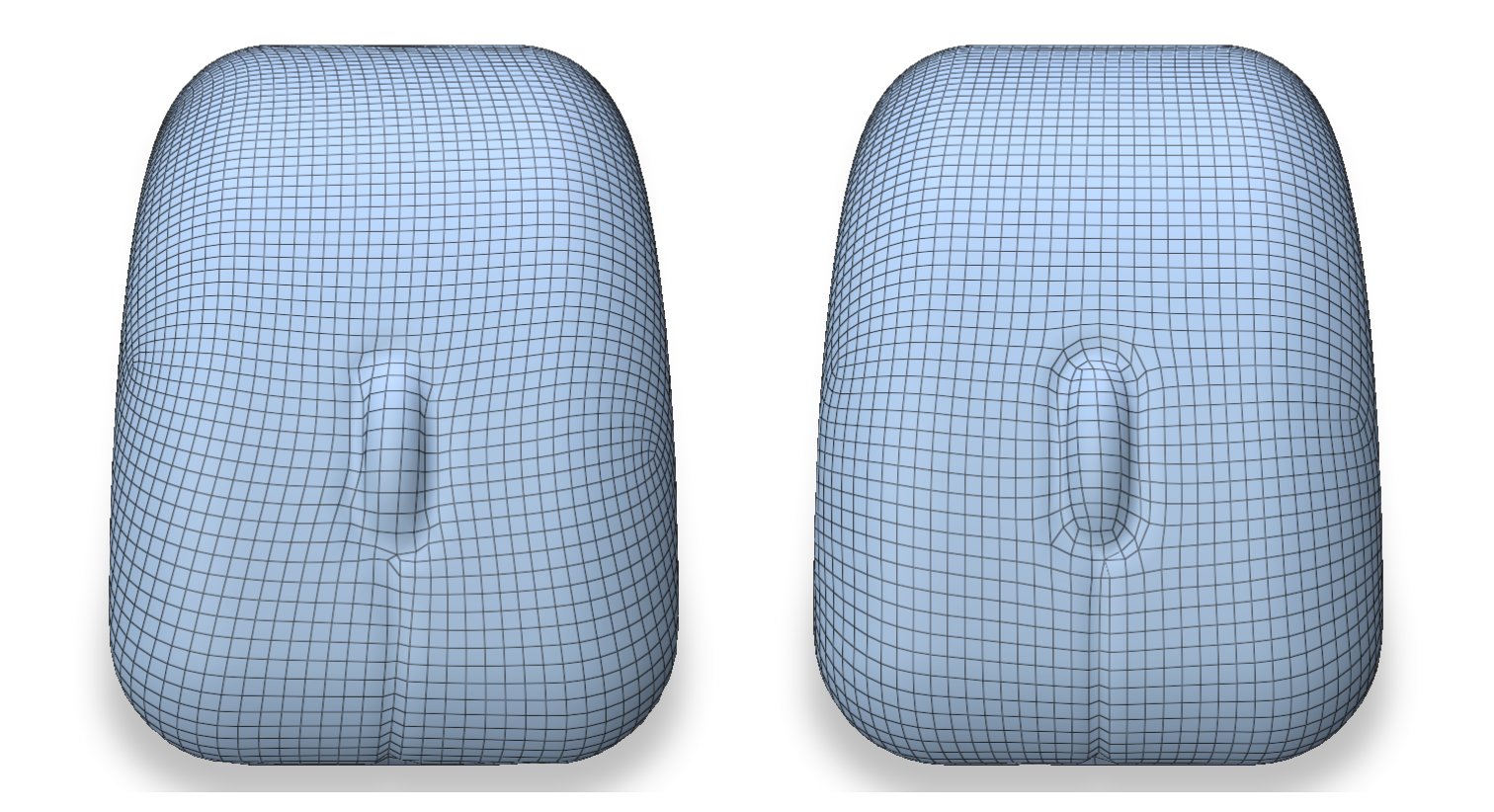}
        \put(20,  \pl){\textcolor{black}{Standard}}
        \put(56.5,  \pl){\textcolor{black}{Sharp Edge Constraints}}
    \end{overpic}
    \vspace{1mm}
    \caption{\textbf{Sharp Feature Alignment}. Our approach can be combined with auxiliary sharp feature constraints to capture subtle geometric details of the surface. While our visual guidance is generally sufficient, users can optionally introduce hard constraints aligned to ``sharp'' edges whose dihedral angle surpasses a user-defined threshold. Using these additional constraints in our solve, we can obtain quads that more closely align to subtle ridges in the geometry such as the curved edges surrounding the scroll wheel on the mouse.}
    \label{fig:sharp-features}
\end{figure}

\smallskip
\noindent\textbf{Sharp Edge Constraints.}\quad
On most meshes, interpolation with our extracted alignment constraints is sufficient to ensure high quality results. However, for some mechanical meshes with sharp edges, using explicit constraints on these edges can improve alignment. We offer a setting where users can detect sharp edges by classifying edges with large dihedral angles and add those edge vectors as hard alignment constraints for the adjacent faces in the interpolations. In \cref{fig:sharp-features}, we show a specific example where using sharp edge constraints can improve our method. Our ``standard'' method without sharp edge constraints (left) produces a quad mesh that is aligned globally, but smooths over the high frequency geometry of the ridges surrounding the mouse's scroll wheel. If we add sharp edge constraints to our method (right), the edges surrounding the scroll wheel have sufficiently large dihedral angles ($>35\degree$) and thus are hard constrained to be aligned with its geometry. Our method smoothly adapts the edge flow to incorporate these geometric constraints along the scroll wheel, while still preserving the visually-informed global alignment over the rest of the mouse body.

\smallskip
\noindent\textbf{Using Additional Views.}\quad
While the $6$ canonical views used by our standard method cover all sides of the mesh, sometimes specific semantic features can be seen more clearly from specific views. To address this, we optionally allow users to generate additional views (beyond the $6$ canonical views) sampled from a Fibonacci distribution around the mesh. This provides additional alignment signal that can help produce better alignment to specific features.

\smallskip
\noindent\textbf{View RANSAC for Improved Multi-View Consistency.}\quad
When running our method on many views as described above, we sometimes find that the results are worsened by a small number of views whose generated images do not agree with the remaining views. To address this concern, we apply a RANSAC algorithm to filter for inlier views that are all in agreement.

In this RANSAC algorithm, we perform a fixed number of iterations. In each iteration, we begin by selecting a random subset of views of a fixed size. We then run our interpolation using only the crosses from view-consistent images for this subset of views. The resulting cross field from this solve acts as our "test model" for these views. We use this test model to filter for inlier views by calculating for each view the average angle difference between the view's initial crosses and the test model's resulting crosses. We filter out any views whose average angle difference is above a set threshold, and the remaining views are the inlier views for this subset. 

If the count of inlier views is too low, we consider the initial subset as a bad selection and move on to the next iteration. Otherwise, we have sufficiently many inlier views for the current subset, so we run the final solve again on only these inlier views to produce the final inlier cross field for this iteration. We then compute the final error of this cross field by computing the average per inlier view error as before, and then averaging the resulting errors across all inlier views. After performing all iterations, we return the final inlier cross field for the iteration that had the lowest final error. 

\section{Visualizations and Implementation Details}
\label{sec:implementation-details}

\begin{figure*}[t]
    \centering
    \includegraphics[width=\linewidth]{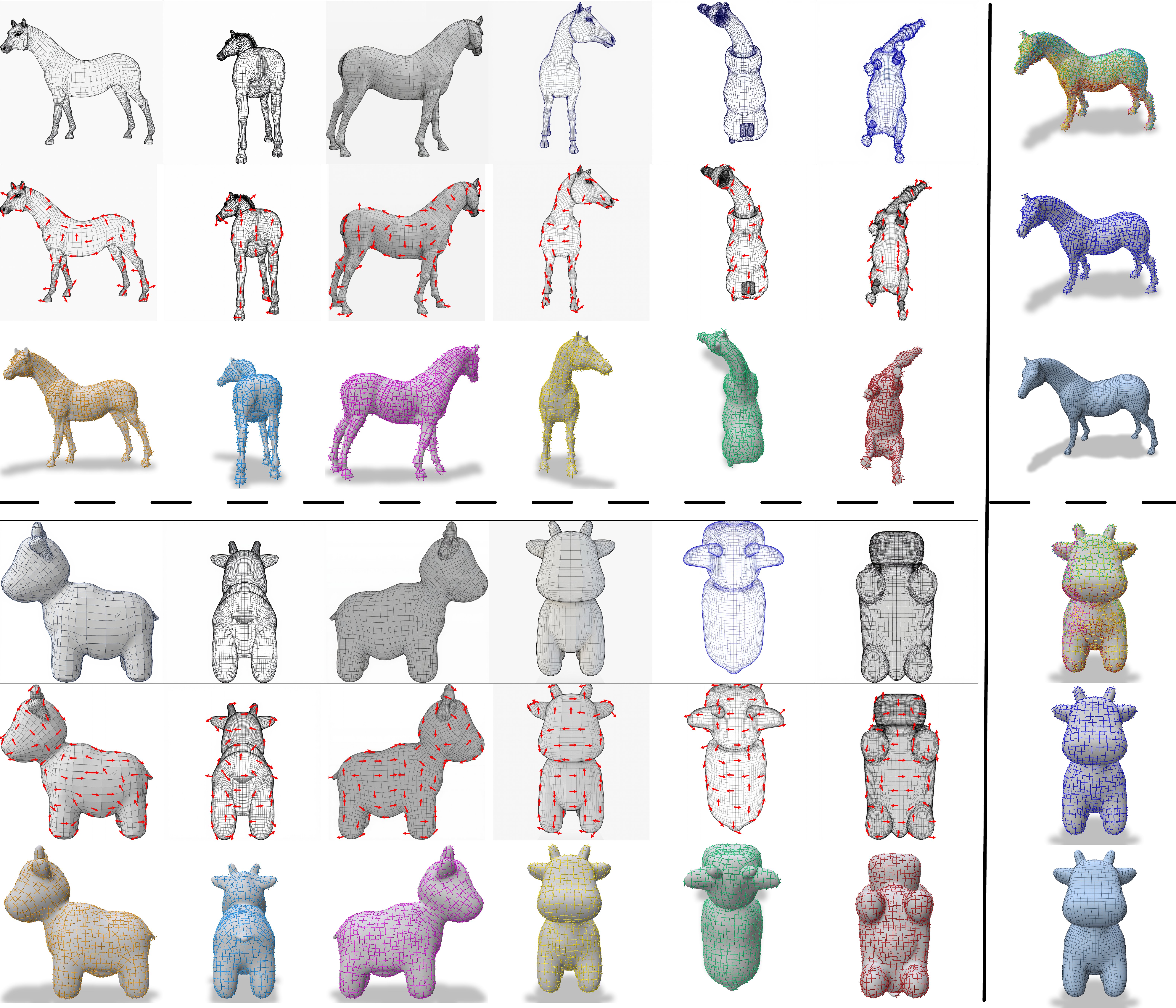}
    \vspace{-4mm}
    \caption{\textbf{Multi-view Alignment Extraction and Cross Field Interpolation}. For both the horse mesh and Spot, we show the generated guidance images in 2D (left, top), a visualization of our extracted gradient directions overlaid on a grayscale version of the guidance images (left, middle), and the extracted and densified cross fields we obtain from each view (left, bottom). We then show the fields from all views composited from a single view (right, top), followed by the single multi-view consistent cross field (right, middle), and finally the quad mesh we extract using this field (right, bottom).}
    \label{fig:all-steps}
\end{figure*}

\noindent\textbf{Visualization of Multi-View Extraction and Interpolation.}\quad
In \cref{fig:all-steps}, we show visualizations of the different intermediate steps in our method's pipeline for two meshes: the horse and Spot. For each mesh, we first show the generated guidance images (left, top) produced by \cref{subsec:generation}. These images often have varying colors and quad scales and can contain many noisy artifacts. From these guidance images, we use the techniques presented in \cref{subsec:grad-extract} to get sparse per-pixel alignment directions at the quad edge lines for each image. We visualize a random subset of these directions in red overlaid on top of the guidance images in the second row (left, middle). In the third row (left, bottom) we show the per-face cross field obtained by performing our first stage interpolation solve on these extracted and back-projected directions for each view (as described in \cref{subsec:per-view-crosses}).

Moving over to the right side of the figure, we overlay the resulting per-view cross fields with crosses colored according to the view they belong to (right, top). Below (right, middle), we show our final cross field in blue after performing our second stage interpolation solve (as described in \cref{subsec:multi-view-crosses}). Lastly, we render the final quad meshing extracted from our cross field (right, bottom).

\begin{figure}
    \centering
    \newcommand{\pl}{-4}
    \begin{overpic}[width=\linewidth]{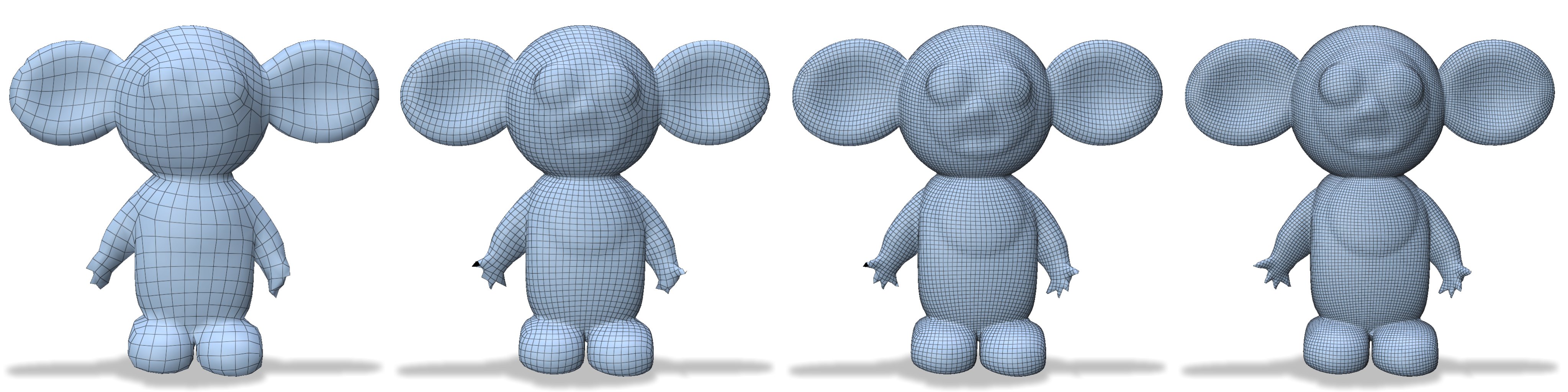}
    \put(6,  \pl){\textcolor{black}{$\text{Scale} = 8$}}
    \put(30,  \pl){\textcolor{black}{$\text{Scale} = 16$}}
    \put(55,  \pl){\textcolor{black}{$\text{Scale} = 24$}}
    \put(80,  \pl){\textcolor{black}{$\text{Scale} = 32$}}
    \end{overpic}
    \vspace{1mm}
    \caption{\textbf{Quad Extraction Scale.} We show various quad sizes for the quad meshes extracted from our cross field. For each quad mesh extraction, we determine the quad size dynamically using a scaled root of the ratio of the original mesh curvature to root area. At lower scales, subtle geometric detail can be missed (missing fingers at $\text{Scale} = 8 \text{ and }16$). At extremely large scales, meshes require many quads which can be computationally expensive to work with ($\text{Scale} = 32$).}%
    \label{fig:quad-size}
\end{figure}
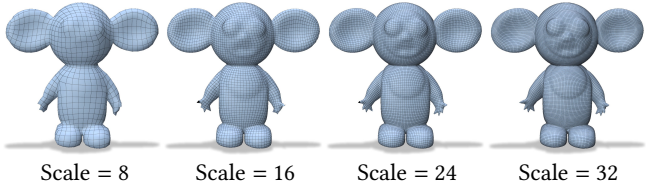

\newpage
\noindent\textbf{Quad Extraction Scale.}\quad
The resolution at which quads are extracted from a cross field can have significant impact on the quality of the quad mesh. If the extraction resolution is too low, subtle geometric features can be smoothed out. If the extraction resolution is too high, quads are added unnecessarily and the mesh becomes computationally expensive to work with. For the results shown in the paper we opt for a dynamic quad scale that is a function of the mesh curvature and total mesh surface area, and a user-defined scale parameter. Specifically, we pass a gradient size to the MiQ parameterization:
\begin{equation}
    gradientSize = s \sqrt{\frac{k}{\sqrt{A}}}
\end{equation}
where $s$ is the user-defined scale factor, $k$ is the curvature given by $\sum_{e} (len(e) \cdot |\theta(e)|)$ (the product of the edge length and dihedral angle integrated over all non-boundary edges), and $A$ is the total surface area of the mesh. In \cref{fig:quad-size}, we visualize quad meshes extracted with different settings for the scale parameter $s$. While we use a single value of $s=16$ for all results shown in the paper, we find that manually tuning $s$ for a given mesh can produce more desirable results.

\smallskip
\noindent\textbf{Prompts Used for 2D the Generative Models.}\quad
For the main results shown in the paper, we use the open source model Flux~\cite{labs2025flux1kontextflowmatching} conditioned on the depth map using a ControlNet~\cite{zhang2023adding} architecture as specified in \cref{subsec:generation}. For this model, we use the following prompt:
\begin{promptbox}
A 3D render of an object with a wireframe quad mesh, equally sized quads, large quads, straight lines, only quads, evenly spaced quadrilateral lines, anti-aliased lines, sharp, no shading, technical CAD illustration, white background.
\end{promptbox}
For specific experiments and figures noted in the paper, we use additional 2D generative models for guidance such as Gemini 3.1 Pro~\cite{team2023gemini} and ChatGPT 5.2~\cite{singh2025openai}. For these two models, we do not have direct access to an explicit ControlNet variant so we encourage better adherence to the geometry by conditioning on not just depth maps, but also normal images and untextured renders as well. To incoprate this additional input signal, we use this prompt:
\begin{promptbox}
Given this untextured render, depth map, and normal map, give me an image from the exact same camera view but with a wireframe quad mesh. Make the quad grid lines black, the mesh light gray, and the background white.
\end{promptbox}
When using Gemini 3.1 Pro, we also optionally allow for generating all $6$ canonical views in a single image grid to give improved multi-view consistency. In this case, we explicitly give instructions to the model about what each portion of the grid image contains and how it relates to the grid image as a whole, \ie:
\begin{promptbox}
Each input image is a 3x2 grid containing 6 views of a 3D object.
...
TEXTURE ALL 6 VIEWS - NO EXCEPTIONS:
    - The Left view (top-left) MUST be fully textured with a wireframe quad mesh
    - The Front view (top-right) MUST be fully textured with a wireframe quad mesh
    - The Right view (middle-left) MUST be fully textured with a wireframe quad mesh
    - The Back view (middle-right) MUST be fully textured with a wireframe quad mesh
    - The Top view (bottom-left) MUST be fully textured with a wireframe quad mesh
    - The Bottom view (bottom-right) MUST be fully textured with a wireframe quad mesh
    - DO NOT leave any view untextured, or partially textured
    - Every single view must show the wireframe quad mesh texture
...
VISUAL CONSISTENCY:
    - The Left view (top-left) and Right view (middle-left) must show opposite sides of the same object with consistent texturing
    - The Front view (top-right) and Back view (middle-right) must show opposite sides with consistent texturing
    - The Top view (bottom-left) and Bottom view (bottom-right) must show opposite sides with consistent texturing
    - Surface details, patterns, and colors must align logically between adjacent views
    - The object should appear as the same 3D object viewed from different angles, not as different objects
\end{promptbox}

\end{document}